\documentclass[preprint,twocolumn]{aastex631}
\usepackage{bm}
\usepackage{threeparttable}
\usepackage{amsmath}
\usepackage{graphicx}
\hypersetup{
  colorlinks=true,
  urlcolor=blue,
}
\usepackage{mhchem} 
\usepackage{chemformula}

\usepackage{makecell}
\usepackage{upgreek}

\begin{document}

\title{Dynamically Coupled Three-dimensional Cloud–Chemistry in Rapidly Rotating Irradiated Atmospheres: an Application to the Compact White Dwarf-Brown Dwarf system ZTFJ0038+2030}

\author[0000-0001-9700-9121]{Yanhong Lai}
\affiliation{Tsung-Dao Lee Institute, Shanghai Jiao Tong University, 1 Lisuo Road, Shanghai 200127, People’s Republic of China}
\email{yanhonglai@sjtu.edu.cn}

\author[0000-0003-2278-6932]{Xianyu Tan}
\affiliation{Tsung-Dao Lee Institute, Shanghai Jiao Tong University, 1 Lisuo Road, Shanghai 200127, People’s Republic of China}
\affiliation{School of Physics and Astronomy, Shanghai Jiao Tong University, 800 Dongchuan Road, Shanghai 200240, People’s Republic of China}
\affiliation{State Key Laboratory of Dark Matter Physics, Shanghai Jiao Tong University, 1 Lisuo Road, Shanghai 200127, People’s Republic of China}
\email{xianyut@sjtu.edu.cn}
\author[0009-0005-9531-7733]{Daphne Broski-Laing}
\affiliation{Department of Astronomy, University of Virginia, Charlottesville, VA 22904, USA}
\author[0000-0003-2969-6040]{Yifan Zhou}
\affiliation{Department of Astronomy, University of Virginia, Charlottesville, VA 22904, USA}

\author[0000-0003-2478-0120]{Sarah L. Casewell}
\affiliation{School of Physics and Astronomy, University of Leicester, University road, Leicester, LE1 7RH, UK}

\author[0000-0003-0825-4876]{Jenni R. French}
\affiliation{School of Physics and Astronomy, University of Birmingham, Birmingham B15 2TT, UK}

\author[0000-0002-3052-7116]{Elspeth K.H. Lee}
\affiliation{Center for Space and Habitability, University of Bern, Gesellschaftsstrasse 6, CH-3012 Bern, Switzerland}

\author[0000-0003-3667-8633]{Joshua D. Lothringer}
\affiliation{Space Telescope Science Institute, Baltimore, MD 21218, USA}

\author[0000-0001-9521-6258]{Vivien Parmentier}
\affiliation{Université de la Côte d'Azur, Observatoire de la Côte d'Azur, CNRS, Laboratoire Lagrange, France}

\author[0000-0002-8163-4608]{Shang-Min Tsai}
\affiliation{Institute of Astronomy \& Astrophysics, Academia Sinica, Taipei 10617, Taiwan}

\begin{abstract}
Clouds and disequilibrium chemistry shape the observable properties of substellar and planetary atmospheres, and JWST time-resolved spectroscopy now enables direct tests of their three-dimensional (3D) structures. 
We present an update to the 3D atmospheric framework \texttt{ADAM}, coupling the dynamical core to a miniature kinetic chemistry scheme (mini-chem) and a saturation-relaxation tracer cloud module. Using this framework, we investigate the atmosphere of the brown dwarf orbiting the white dwarf ZTFJ0038+2030 in a 10-hour orbit.
The brown dwarf receives stellar irradiation comparable to its internal heat flux, placing the atmosphere in a regime where the circulation is characterized by a standing Matsuno--Gill wave response and equatorial superrotation. Rapid rotation confines these structures closer to the equator, producing more symmetric broadband light curves with smaller hotspot offsets than canonical hot Jupiters, in agreement with recent JWST observations of this system. Circulation-driven disequilibrium chemistry enhances global-mean CO and CO$_2$ while slightly reducing H$_2$O, CH$_4$, and NH$_3$ relative to chemical equilibrium. 
Although most species are globally homogenized, rapid rotation limits global-scale horizontal mixing and allows circulation-driven vertical transport to maintain strong CO and CO$_2$ inhomogeneities in the photosphere. These chemical inhomogeneities produce asymmetric phase curves in the CO and CO$_2$ absorption bands, which could qualitatively explain the observed CO$_2$ light-curve asymmetry.
When clouds are included, Na$_2$S clouds dominate the photosphere and weaken the large-scale circulation, lowering photospheric CO and CO$_2$ abundances, while MgSiO$_3$ and Fe clouds rain out at deeper levels. Because the dominant Na$_2$S clouds reside mostly above the quench region, cloud-driven convection has limited impact on the quenched abundances, which are instead regulated mainly by large-scale circulation.
\end{abstract} 


\section{Introduction}
\label{sec:intro}

\subsection{Compact WD-BD systems as a unique probe of planetary circulation regimes}

\begin{figure*}
    \centering
    \includegraphics[width=0.95\textwidth]{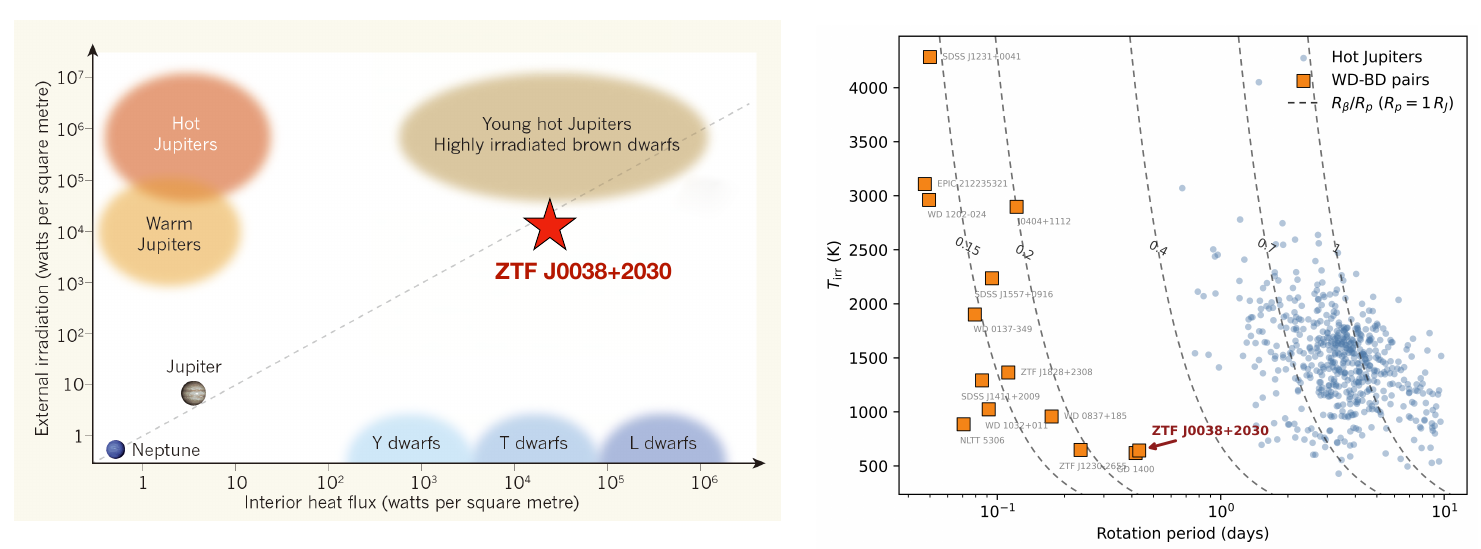}
    \caption{\textit{Left}: a regime diagram of internal heat flux and stellar irradiation for different classes of fluid planetary bodies from \cite{showman2016}. ZTFJ0038+2030 (red star) lies in a regime where internal heat and external irradiation are both intermediately strong and comparable. 
    \textit{Right}: rotation period and irradiation temperature ($T_{\rm irr}$) for confirmed transiting hot Jupiters and compact BD--WD systems, with ZTFJ0038+2030 highlighted by a red arrow. Hot Jupiters are selected from the NASA Exoplanet Archive as planets with radii $R_p > $8 R$_{\oplus}$ and orbital periods $P < 10$ days, assuming synchronous rotation. Gray dashed curves show contours of $R_{\beta}/R_p$, where $R_{\beta}=\sqrt{c/\beta}$ is the equatorial deformation radius and $R_p=1\,$R$_\text{J}$ is assumed. Here $\beta$ is the latitudinal gradient of the Coriolis parameter, and $c\sim2NH$ is the horizontal phase speed of long-vertical-wavelength gravity waves \citep{tan2020}, estimated for isothermal atmospheres with temperature $T_{\rm irr}$. $N$ and $H$ are the Brunt--Väisälä frequency and scale height, respectively.
    Data are from the NASA Exoplanet Archive (\url{https://exoplanetarchive.ipac.caltech.edu}).
    }
    \label{fig:planet}
\end{figure*}

Atmospheric circulation in gaseous planets and substellar objects is powered by two primary energy sources: external stellar irradiation and internal heat flux. The relative importance of these sources places objects in distinct regimes \citep{showman2016}. As illustrated in the left panel of Figure~\ref{fig:planet}, the Solar-System gaseous planets occupy a low-flux regime in which stellar and internal heating are both modest and broadly comparable. Their long radiative timescales favor persistent large-scale structures, including stable zonal jets and long-lived vortices \citep{showman2010}. The two best-observed populations of extrasolar gaseous worlds occupy the opposite extremes of parameter space. Hot Jupiters (HJs), assumed to be tidally locked, reside in a strongly irradiated regime, where intense and stationary day-night forcing drives immense global-scale circulation characterized by standing waves, broad superrotating equatorial jets, and overturning circulation. In contrast, isolated brown dwarfs (BDs), as well as young directly imaged giant planets, are predominantly powered by internal heat flux and commonly exhibit transient waves, turbulence and vortical structures with sizes typically smaller than their radii. A recent review of these circulation regimes can be found in \citet{showman2020review}.

BDs in compact white dwarf-brown dwarf (WD-BD) systems occupy a comparatively unexplored region of parameter space in which both stellar irradiation and internal heat flux are strong and can be of comparable magnitude. As a result, they provide a natural laboratory for investigating how atmospheric circulation responds to the simultaneous action of stellar irradiation and internal heat flux, two energy sources that, when acting separately, produce markedly different dynamical behaviors. Understanding this intermediate regime is important for connecting the atmospheric dynamics of strongly irradiated planets and self-luminous substellar and planetary objects.

A small but growing population of compact WD-BD systems with measured orbital periods has been identified (squares in the right panel of Figure~\ref{fig:planet}): GD~1400 \citep{farihi2004}, WD~0137$-$349 \citep{burleigh2006}, NLTT~5306 \citep{steele2013}, WD~0837+185 \citep{casewell2012}, SDSS~J1411+2009 \citep{beuermann2013}, SDSS~J1557+0916 \citep{farihi2017}, WD~1202$-$024/SDSS~J1205$-$0242 \citep{parsons2017}, SDSS~J1231+0041 \citep{parsons2017}, EPIC~212235321 \citep{casewell2018}, WD~1032+011 \citep{casewell2020}, ZTF~J0038+2030 \citep{van2021}, ZTF~J1828+2308 \citep{parsons2025}, J0404+1112 \citep{dewit2025}, and ZTF~J1230$-$2655 \citep{parsons2025}. These systems are generally believed to undergo a period of post-common-envelope evolution, during which the BD companion becomes engulfed within the envelope of the giant progenitor and spirals inward while helping to eject the envelope \citep{politano2004}. The resulting compact configurations lead to intense irradiation and rapid synchronous rotation of the BD companion.

In addition to their intermediate heating regime, rotation constitutes a fundamental control on planetary circulation through the Coriolis force, which sets the characteristic scales of jets, waves, and vortices \citep[e.g.,][]{vallis2017atmospheric}. For tidally locked bodies with stationary day-night forcing, a useful measure of rotational effect is $R_\beta/R_p$, similar to the weak-temperature-gradient (WTG) parameter \citep{pierrehumbert2019}, where $R_\beta$ is the equatorial deformation radius and $R_p$ is the object radius. The right panel of Figure~\ref{fig:planet} compares irradiation temperatures and rotation periods for transiting HJs and known WD-BD systems. Whereas HJs typically rotate synchronously with periods of several days, BDs in compact WD-BD systems are synchronized on timescales of only a few hours. Such rapid rotation substantially reduces the equatorial deformation radius and strengthens rotational constraints on atmospheric motions and heat transport. As shown by the contours in Figure~\ref{fig:planet}, compact WD-BD systems and HJs occupy distinct regions in $R_\beta/R_p$, suggesting fundamentally different regimes of atmospheric circulation and heat transport \citep{tan2020}. 

Recent HST/WFC3 light-curve observations of multiple compact WD–BD pairs typically show negligible phase offsets, with the peak thermal emission from the BD occurring close to the phase of maximum stellar heating 
\citep{Zhou2022,Lew2022,Amaro2023,Amaro2024,Lothringer2024,French2024,Amaro2025}. These findings are consistent with the modeling studies of \cite{tan2020} and \cite{Lee2020wdbd}, which demonstrated that as $R_\beta/R_p$ decreases to values characteristic of the rapidly rotating WD–BD regime, the meridional extents of the equatorial standing wave and jet substantially decrease, resulting in more symmetric light curves and reduced day-to-night heat transport of the irradiative energy. Note that other irradiated BDs with larger $R_\beta/R_p$, such as KELT-1b orbiting a main-sequence star, can still exhibit appreciable hotspot offsets \citep{beatty2019}.
However, the implications of this circulation regime of $R_\beta/R_p\ll 1$ for cloud formation and chemical mixing remain unclear--a question that becomes increasingly important in the JWST era, where broad-wavelength coverage (0.6--5.3~$\upmu$m) enables detailed characterization of the 3D atmospheres of BDs in WD–BD systems \citep{zhou2024jwst}.

\subsection{The importance of clouds and disequilibrium chemistry}
Clouds and disequilibrium chemistry are ubiquitous in planetary and substellar atmospheres and play central roles in shaping their observable properties \citep{marley2015}. Clouds are commonly inferred in exoplanet atmospheres through muted or featureless transmission and thermal emission spectra \citep[e.g.,][]{demory2013,knutson2014,sing2016,feinstein2023}. In BDs, spatially inhomogeneous cloud structures are strongly supported by infrared light-curve variability, commonly attributed to rotational modulation by patchy cloud coverage \citep[e.g.,][]{artigau2009,radigan2014,buenzli2015,metchev2015,biller2017}. Direct spectroscopic evidence of silicate grains has been obtained for both substellar and transiting planetary atmospheres \citep[e.g.,][]{cushing2006,suarez2023,grant2023}.
Observational evidence further indicates that many planetary and substellar atmospheres depart substantially from thermochemical equilibrium owing to atmospheric transport and photochemistry \citep[e.g.,][]{saumon2000,sorahana2012,blumenthal2018,miles2020,leggett2021,miles2023,tsai2023,bell2024,sing2024}. Such disequilibrium chemistry alters the abundances of key absorbing species, leading to deviations from chemical-equilibrium spectral predictions and modifying observed atmospheric signatures.

Large-scale circulation, convection, and wave-driven transport redistribute condensable material and chemically active species both vertically and horizontally \citep{ackerman2001,parmentier2013,bordwell2018,gao2020,drummond2020,tan2022bd,parmentier2026}, thereby linking cloud formation and disequilibrium chemistry. These interactions are inherently two-way. Cloud radiative feedback modifies atmospheric thermal structures and circulation \citep{tan2021a,tan2021b,lee2024,tan2025,teinturier2026}, thereby altering chemical transport, while localized cloud heating can trigger convection and enhance vertical mixing \citep{tan2019}. Disequilibrium chemistry can likewise modify atmospheric opacity and radiative heating \citep{mukherjee2025,agundez2025}. Clouds, chemistry, radiation, and dynamics therefore form a tightly coupled system that must be treated self-consistently.

Capturing these interactions requires 3D models that couple atmospheric dynamics, radiation, cloud formation, and kinetic chemistry. However, fully coupled frameworks remain uncommon. Most 3D studies that include cloud radiative feedback assume chemical equilibrium \citep[e.g.,][]{roman2019,roman2021,tan2021a,tan2021b,parmentier2021,Teinturier2024,tan2025}, whereas models focused on disequilibrium chemistry often neglect clouds \citep[e.g.,][]{bordwell2018,steinrueck2019,drummond2020,lee2023,lee2023dynamically}. Other approaches treat clouds and chemical species as passive tracers, thereby omitting their feedback on radiation and circulation \citep{parmentier2013,komacek2019mixing,tan2022bd}.
Such self-consistent 3D models are also essential for interpreting retrieval results by linking retrieved atmospheric properties to the underlying circulation, transport processes, and cloud--chemistry interactions \citep{madhusudhan2018,barstow2020}.

A notable exception is the recent work of \citet{lee2024}, who coupled the general circulation model with kinetic chemistry and a tracer-based cloud formation scheme for directly imaged sub-Jovian exoplanets. Their results revealed strong feedback among cloud formation, atmospheric dynamics, and chemical abundances, demonstrating the importance of a fully self-consistent treatment. That study focused on weakly irradiated, directly imaged planets with negligible external forcing. 
Compact WD--BD systems instead experience strong day--night forcing, substantial internal heat flux, and rapid rotation, placing them in a distinct dynamical regime. The coupled behavior of clouds, dynamics, and disequilibrium chemistry under these conditions remains largely unexplored.

\subsection{ZTFJ0038+2030}
JWST began obtaining spectroscopic light curves of several WD-BD systems, covering 1 to 5 microns through program JWST-GO-4967 \citep{zhou2024jwst}. Among these and other known WD-BD systems, ZTFJ0038+2030B (hereafter ZTF0038B) lies between typical HJs and the most rapidly rotating WD-BD systems in the $R_\beta/R_p$ space. It therefore samples an intermediate regime in which rotation exerts a moderate influence on the atmospheric circulation, while receiving external irradiation comparable to its internal heat flux (Figure \ref{fig:planet}). As a result of intermediate internal and external energy fluxes, ZTF0038B has moderate atmospheric temperatures, with nightside and dayside effective temperatures inferred from the joint retrieval of JWST phase-resolved spectra of $\sim$922 and $\sim$982~K, respectively \citep{broskilaing2026}. This places ZTF0038B closer to typical warm Jupiters that are subject to disequilibrium chemistry driven by atmospheric quenching \citep[e.g.,][]{fortney2020}. Therefore, ZTF0038B provides a unique opportunity for examining how mixed thermal forcing and rapid rotation jointly regulate atmospheric circulation, heat redistribution, cloud formation, and chemical transport. This system and the JWST spectroscopic light-curve are described and analyzed in \cite{broskilaing2026}. 

In this work, we investigate the 3D coupling between atmospheric circulation, cloud formation, and disequilibrium chemistry in compact WD--BD systems using an updated 3D framework, \texttt{ADAM} (the ADvanced Atmospheric MITgcm), with ZTF0038B as a reference case. 
The paper is organized as follows. Section~\ref{sec:model} describes the model setup and experimental design. Section~\ref{sec:ztf_results} presents the circulation, cloud, and chemical structures in the ZTF0038B atmosphere. Section~\ref{sec:ztf_spectra} presents the observational signatures of ZTF0038B. Section~\ref{sec:conclude} summarizes our main findings.

\section{Model Setup and Experimental design}\label{sec:model}

\subsection{GCM modelling using \texttt{ADAM}}\label{sec:gcm}

We model the atmosphere using an updated 3D framework, \texttt{ADAM}, which couples the dynamical core with the SPARC radiative transfer module, active cloud tracers, and passive chemical tracers \citep{showman2009,mehta2025}. \texttt{ADAM} with SPARC (formerly known as SPARC/MITgcm) is built upon the MITgcm dynamical core \citep{adcroft2004} and a non-gray radiative transfer model developed by \cite{marley1999}. This framework has been widely applied to the atmospheric modeling of isolated BDs and HJs \citep[e.g.,][]{showman2009,kataria2013,parmentier2013,komacek2017,tan2019uhj,tan2021a,komacek2022cloud,steinrueck2023,mehta2025}, as well as warm mini-Neptunes \citep{lai2026}.

The dynamical core solves the primitive equations on a cubed sphere grid in pressure coordinates,  assuming local hydrostatic balance \citep{adcroft2004,Vallis2006}. The validity of the hydrostatic approximation on the global circulation can be evaluated from the square of the aspect ratio $(H/L)^2$, where $H$ and $L$ are representative vertical and horizontal length scales of the circulation (\citealt{Vallis2006}, chapter 2.7). For ZTF0038B, the characteristic horizontal and vertical scales are approximately 10$^7$~m and 10$^3$~m, respectively, giving a small value of $(H/L)^2\sim10^{-8}$ (Table~\ref{tab:paras}). We therefore expect the hydrostatic framework to properly capture the large-scale wave dynamics and jets that are central to our analysis, while small-scale non-hydrostatic motions are not explicitly resolved. 

The deep layers in the modeled domain can be convectively unstable. In addition, cloud radiative effects can trigger local convective instability in regions with sufficiently thick clouds. Such rapid convection and the mixing of entropy and tracers are not explicitly resolved in global models but can only be parameterized. We represent the effects of sub-grid convective mixing of specific entropy and tracers (including condensable vapor, condensates, and chemical species) using a diffusion process based on the classical mixing length theory \citep{Gierasch1968}. The time- and spatially varying vertical diffusion coefficient $K_{zz}$ is given by
\begin{equation}
K_{\rm zz} =
\begin{cases}
0, & \text{if } \frac{\partial \ln{T}}{\partial \ln{p}} \le (\frac{\partial \ln{T}}{\partial \ln{p}})_{\rm ad} \\
L^2 \sqrt{\frac{g}{H} [\frac{\partial \ln{T}}{\partial \ln{p}}-(\frac{\partial \ln{T}}{\partial \ln{p}})_{\rm ad}]} , & \text{else }\\
\end{cases}
\label{eq:kzz}
\end{equation}
where $T$ and $p$ are atmospheric temperature and pressure, $g$ is gravity, $H=RT/g$ is the atmospheric scale height, $R$ is the specific gas constant, and $L=\alpha H$ is the characteristic mixing length, fixed to one scale height ($\alpha=1$). Convective diffusion is effective only in convectively unstable regions, where the temperature lapse rate exceeds the adiabatic value, i.e., $\frac{\partial \ln{T}}{\partial \ln{p}} > (\frac{\partial \ln{T}}{\partial \ln{p}})_{\rm ad}= \frac{R}{c_p}$, where $c_p$ is the heat capacity. Outside convective regions, a small background diffusion coefficient of K$_{\rm zz, min}$ = $1~\rm m^2\,s^{-1}$ is imposed to crudely represent weak mixing in stably stratified layers and to ensure numerical stability \citep{tan2019}. Such a small value does not significantly influence the large-scale circulation \citep{tan2019}.

To represent momentum exchange between the deep interior and the weather layer, we apply a Rayleigh drag at pressures greater than 300 bar, following the formulation in \cite{komacek2016}, relaxing horizontal winds toward zero with a prescribed timescale of $10^6$ s in the bottom layer. A fourth-order Shapiro filter is applied to the temperature field to maintain numerical stability \citep{Shapiro1970}.

Radiative heating and cooling rates are computed using the non-gray, two-stream radiative transfer scheme of \citet{marley1999}, based on the correlated-$k$ method. Chemical equilibrium is assumed when computing gas opacities in the radiative transfer module; thus, the thermal structure does not respond to disequilibrium chemical abundances. Previous studies have shown that disequilibrium chemistry typically modifies atmospheric temperatures by less than $\sim100$ K for solar or supersolar metallicity atmospheres \citep[e.g.,][]{drummond2016,steinrueck2019,mukherjee2024,mukherjee2025,agundez2025}. For ZTF0038B, the largest disequilibrium changes occur in CO and CO$_2$, whereas H$_2$O and CH$_4$ vary more modestly (Section~\ref{sec:ztf_chem}). We therefore expect the associated opacity feedback to be strongest in specific CO- and CO$_2$-sensitive bands, although local thermal effects may still be important. A fully coupled treatment of disequilibrium chemistry and radiative transfer is required to quantify these effects. 


\begin{table}[!t]
\caption{Adopted GCM simulation parameters for ZTF0038B. \label{tab:paras}}
\centering
\begin{tabular}{ll}
\hline
Parameter & Value \\
\hline
WD temperature ($T_\star$) & 10900 K \\
WD mass ($M_\star$) & 0.505 M$_{\odot}$ \\
WD radius ($R_\star$) & 0.01429 R$_{\odot}$ \\
BD mass ($M$) & 0.0593 M$_{\odot}$ \\
BD radius ($R_{\rm bd}$) & 0.0783 R$_{\odot}$ \\
BD gravity ($g$) & 2660 m\,s$^{-2}$ \\
Irradiation temperature ($T_{\rm irr}$) & 655 K \\
Internal temperature ($T_{\rm int}$) & $\sim 920$ K \\
Rotation period ($P$) & 10.4 hours \\
Atmospheric composition & solar metallicity  \\
Heat capacity ($c_p$) & 13000 J\,kg$^{-1}$\,K$^{-1}$ \\
Specific gas constant ($R$) & 3714 J\,kg$^{-1}$\,K$^{-1}$ \\
Lower boundary pressure ($P_{\rm bot}$) & 500 bar \\
Upper boundary pressure ($P_{\rm top}$) & $5\times10^{-4}$ bar \\
Horizontal resolution & C48 \\
Vertical resolution & 53 layers \\
Dynamical timestep & 10 s \\
Chemical timestep & 100 s \\
Cloud timescale ($\tau_c$) & 100 s \\
$q_{\rm deep}$ for MgSiO$_3$ & $10^{-4}$ kg\,kg$^{-1}$ \\
$q_{\rm deep}$ for Fe & $10^{-4}$ kg\,kg$^{-1}$ \\
$q_{\rm deep}$ for Na$_2$S & $5.9\times10^{-5}$ kg\,kg$^{-1}$ \\
\hline
\end{tabular}
\begin{tablenotes}
\footnotesize
\item Note: M$_{\odot}$ and R$_{\odot}$ represent solar mass and radius. The basic WD and BD parameters of ZTF0038B are taken from \cite{van2021}. 
\end{tablenotes}
\end{table}

\subsection{Cloud formation scheme}\label{sec:cloudscheme}
We adopt the cloud formation scheme of \cite{tan2021a,tan2021b}. For each condensable species, two tracer equations representing the mass mixing ratios of condensable vapor ($q_v$) and cloud particles ($q_c$) are solved alongside the dynamical equations:
\begin{equation}
    \frac{dq_v}{dt}=-\delta \frac{q_v - q_c}{\tau_c} + (1-\delta)\frac{\min(q_s-q_v, q_c)}{\tau_c} + Q_{deep}, 
    \label{eq:qv}
\end{equation}
\begin{equation}
    \frac{dq_c}{dt}=\delta \frac{q_v - q_c}{\tau_c} - (1-\delta)\frac{\min(q_s-q_v, q_c)}{\tau_c} -\frac{\partial (\langle q_c  V_s\rangle)}{\partial p}.
    \label{eq:qc}
\end{equation}
Here, $\tau_c = 100$ s is the conversion timescale between vapor and condensate, $q_s$ is the saturation mixing ratio, $V_s$ is the particle settling velocity (summarized in, e.g., \citealt{ackerman2001,parmentier2013}), and $\langle q_c V_s \rangle$ is the normalization over the particle size distribution. The deep source term
$Q_{\rm deep} = -(q_v-q_{\rm deep})/\tau_{\rm deep}$ is applied only at pressures greater than 200 bar and relaxes the vapor abundances toward a prescribed deep mixing ratio $q_{\rm deep}$ with a timescale $\tau_{\rm deep}=1000$ s. 

Condensation occurs when $q_v>q_s$, in which case $\delta=1$, while evaporation occurs when $q_v<q_s$ with $\delta=0$. The saturation mixing ratio is given by $q_s = P_T q_{\rm deep}/p$, where $P_T$ denotes the temperature-dependent saturation vapor pressure for each species (in unit of bars) \citep{visscher2010, morley2012}. 
Cloud particle sizes follow a lognormal distribution with a reference radius of 1.5 $\upmu$m and a non-dimensional width parameter $\sigma=0.5$. 
The characteristic particle sizes in exoplanet and brown dwarf atmospheres are often not directly measured. The sizes predicted from models are uncertain and depend on vertical mixing, gravity, and the thermochemical environment \citep[e.g.,][]{ackerman2001,gao2018}. Previous microphysical studies of substellar atmospheres suggest effective particle sizes of order of a few $\upmu$m with a broad size distribution under mixing and gravity conditions broadly comparable to those considered here \citep{gao2018,helling2019}. We therefore adopt 1.5 $\upmu$m as a representative micron-sized value for ZTF0038B.

We include three representative condensate species commonly expected in BD atmospheres: MgSiO$_3$, Fe, and Na$_2$S \citep{lodders2002,lodders2006,visscher2010,morley2012}. For Na$_2$S, we assume solar elemental abundances and set the deep mixing ratio to $q_{\rm deep}=5.9\times10^{-5}$ kg\,kg$^{-1}$ \citep{lodders2003}. For MgSiO$_3$ and Fe, the deep mixing ratios are set to $10^{-4}$ kg\,kg$^{-1}$, an order of magnitude smaller than their nominal solar-abundance values, for numerical stability. Test simulations using the full Fe and MgSiO$_3$ abundances became numerically unstable due to strong cloud-radiative feedback. Reducing the cloud abundances while keeping their influences in the dynamics and spectra has been a strategy in other GCM studies \citep[e.g.,][]{Teinturier2024,Kennedy2025}. With retrieved dayside and nightside effective temperatures of order $\sim$900--1000 K \citep{broskilaing2026}, ZTF0038B lies in the T-dwarf temperature regime, where iron and silicate clouds are generally expected to have dissipated from the photosphere \citep{burrows2006,visscher2010,apai2013,radigan2014}. Consistently, Fe and MgSiO$_3$ clouds reside mostly below the observable photosphere because of the high surface gravity of ZTF0038B ($g\approx2660$~m\,s$^{-2}$, roughly 100 times Jupiter’s gravity), limiting their direct radiative effect on the observable atmosphere (Section~\ref{sec:ztf_results}).  

Cloud radiative effects are included through wavelength-dependent absorption and scattering. The extinction coefficient, scattering coefficient, and asymmetry factor are interpolated from precomputed Mie tables \citep{batalha2020}, using refractive indices from \citet{jager2003} for MgSiO$_3$, \citet{palik1997} for Fe, and \citet{montaner1979} and \citet{khachai2009} for Na$_2$S.

\subsection{Disequilibrium Chemistry: mini-chem}\label{sec:mini-chem}
We include time-dependent chemical kinetics by coupling \texttt{ADAM} with mini-chem, an open-source miniature chemical network designed for gas giant atmospheres \citep{tsai2022,lee2023}. Mini-chem employs precomputed net forward reaction rate tables, enabling efficient integration of chemical kinetics with a reduced number of species and reactions evolved in the system. We adopt the NCHO network consisting of 12 chemically active species (OH, H$_2$, H$_2$O, H, CO, CO$_2$, O, CH$_4$, C$_2$H$_2$, NH$_3$, N$_2$, and HCN) and 10 reactions.
This reduced network significantly improves computational efficiency, though it requires assuming a metallicity when constructing the reaction tables. 

The adopted NCHO network includes the major molecular species considered in current retrievals of ZTF0038B, including H$_2$O, CH$_4$, CO, CO$_2$, and NH$_3$ \citep{broskilaing2026}, and therefore captures the principal C--N--O thermochemical pathways relevant to the observable atmosphere. This choice is also broadly consistent with the chemistry of late-T and Y dwarfs, whose observable spectra are primarily shaped by H$_2$O, CH$_4$, and NH$_3$ \citep[e.g.,][]{cushing2011,Kirkpatrick2012}, with vertical mixing producing important disequilibrium abundances of CO \citep{sorahana2012,leggett2012,miles2020}. In hydrogen-helium rich atmospheres, \citet{tsai2022} showed that, mini-chem using a NCHO network agrees with the full VULCAN chemical kinematic calculations reasonably well.


Helium is included as an inert tracer contributing to the total volume mixing ratio (VMR) but not participating in chemical reactions. All chemical tracers are integrated using a chemical timestep of $\tau_{\rm chem}=100$ s, advected by the resolved 3D circulation throughout the atmosphere, and additionally mixed vertically by the parameterized convective diffusion K$_{\rm zz}$ within convectively unstable regions. Chemical species do not feed back onto the radiative heating and cooling calculations (Section~\ref{sec:gcm}), and photochemical processes are not included in mini-chem. Photochemical products may nevertheless contribute important opacity sources in temperate atmospheres with irradiation temperatures below $\sim1000$ K \citep{gao2020}.

\begin{figure}
    \centering
    \includegraphics[width=0.95\linewidth]{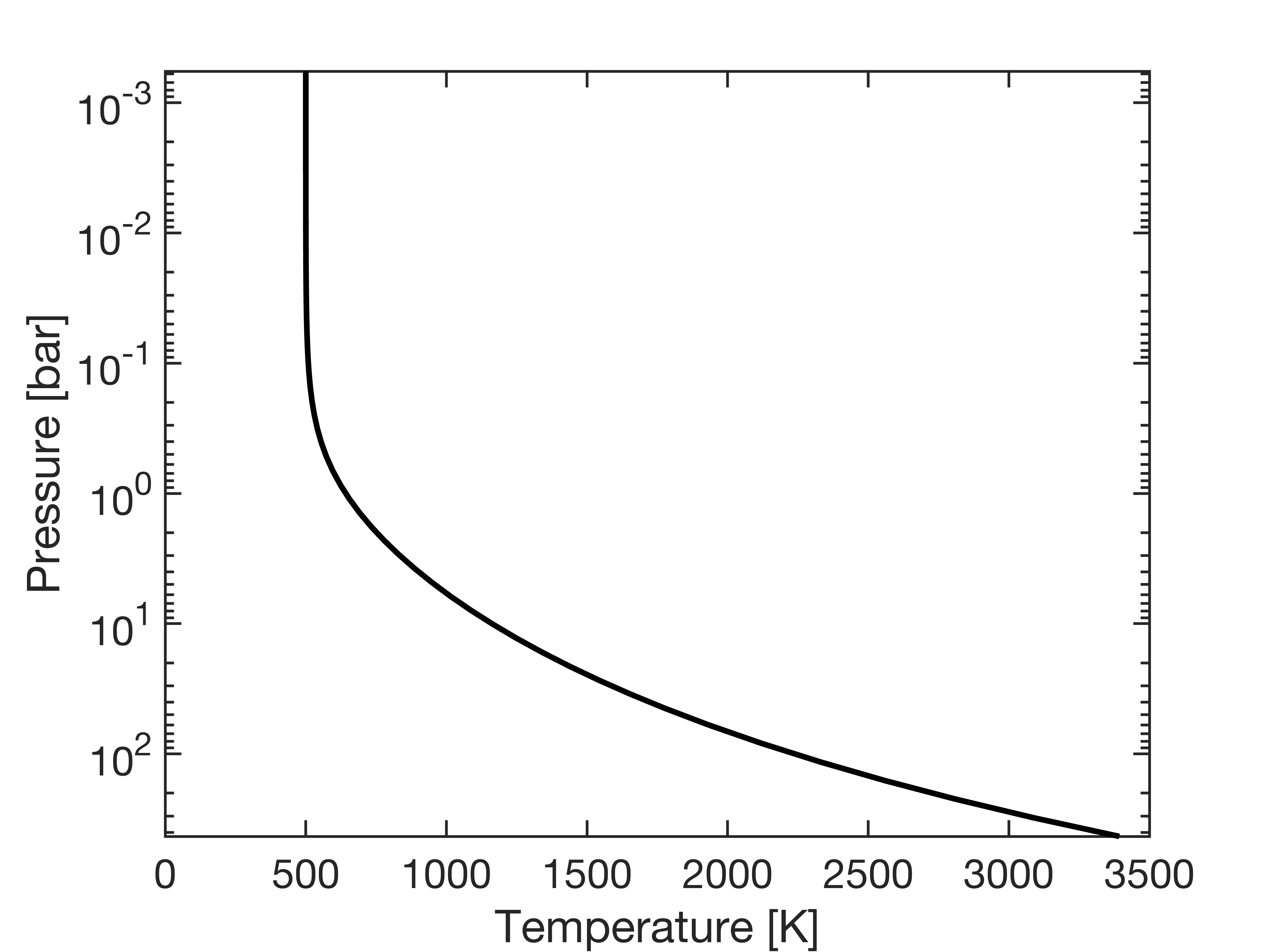}
    \caption{Initial temperature--pressure profile used in both the cloud-free and cloudy simulations.}
    \label{fig:tp}
\end{figure}

\subsection{Radiative transfer post-processing: PICASO}\label{sec:picaso}
To compute observable spectra, we post-process the 3D GCM outputs using PICASO version 3.2.2\footnote{\url{https://github.com/natashabatalha/picaso}}, an open-source radiative transfer code for computing the reflected, thermal, and transmission spectra of planets and brown dwarfs \citep{batalha2019,mukherjee2023}. Following previous applications of PICASO to 3D GCM outputs \citep{adams2022,robbins2022,tan2024}, we calculate thermal emission spectra using the spatially and temporally varying temperature structure, gas abundances, and cloud opacities from the GCM simulations. 

We compute spectra under both chemical equilibrium and disequilibrium conditions. Because PICASO and mini-chem adopt different equilibrium chemistry frameworks, their predicted equilibrium abundances, particularly for H$_2$O, can differ \citep{batalha2019,tsai2017,tsai2022}. To ensure a consistent comparison when evaluating the effects of atmospheric transport, we compute chemical equilibrium spectra using the equilibrium abundances assumed in mini-chem rather than the default PICASO equilibrium chemistry. Disequilibrium spectra are calculated using the chemical abundances obtained from the coupled mini-chem GCM simulations.
Spectra are first calculated using the opacity-resampling technique at a native resolution of $R\sim10{,}000$ over 0.3–14 $\upmu$m and then binned to $R\sim150$. To make the post-processing computationally feasible, the horizontal grid resolution used in PICASO is downgraded to 60$\times$60 of the hemisphere, equivalent to 3$^{\circ}$ per cell, which is sufficient to capture the effect of medium- to large-scale surface inhomogeneity on the spectra. 
Under both assumptions, the local VMRs of the primary absorbing species, H$_2$O, CO, CO$_2$, CH$_4$, C$_2$H$_2$, NH$_3$, and HCN are used to construct the gas opacities.


\subsection{Experimental design}\label{sec:experiments}
ZTF0038B receives external irradiation from its WD companion, corresponding to an irradiation temperature $T_{\rm irr}=655$ K estimated from
\begin{equation}
    T_{\rm irr} = T_\star \sqrt{\frac{R_\star}{2d}},
    \label{eq:Tirr}
\end{equation}
where $T_\star$ and $R_\star$ are the effective temperature and radius of the WD, and $d$ is the semi-major axis \citep{van2021}. 
We adopt a fiducial intrinsic temperature of $T_{\rm int}\approx920$ K, motivated by the BD-only JWST spectrum of ZTF0038B \citep{broskilaing2026}, whose nightside emission is comparable to isolated mid--late T dwarfs \citep{kirkpatrick2021a}. This value is not intended to reproduce the exact internal temperature parameter inferred in the JWST retrieval ($\sim$1240 K), but serves as a physically motivated fiducial internal heat flux. Using the fiducial parameters of ZTF0038B (Table~\ref{tab:paras}), we conduct two simulations: a cloud-free case and a cloudy case including MgSiO$_3$, Fe, and Na$_2$S condensates.

\begin{figure*}
    \centering
    \includegraphics[width=0.95\textwidth]{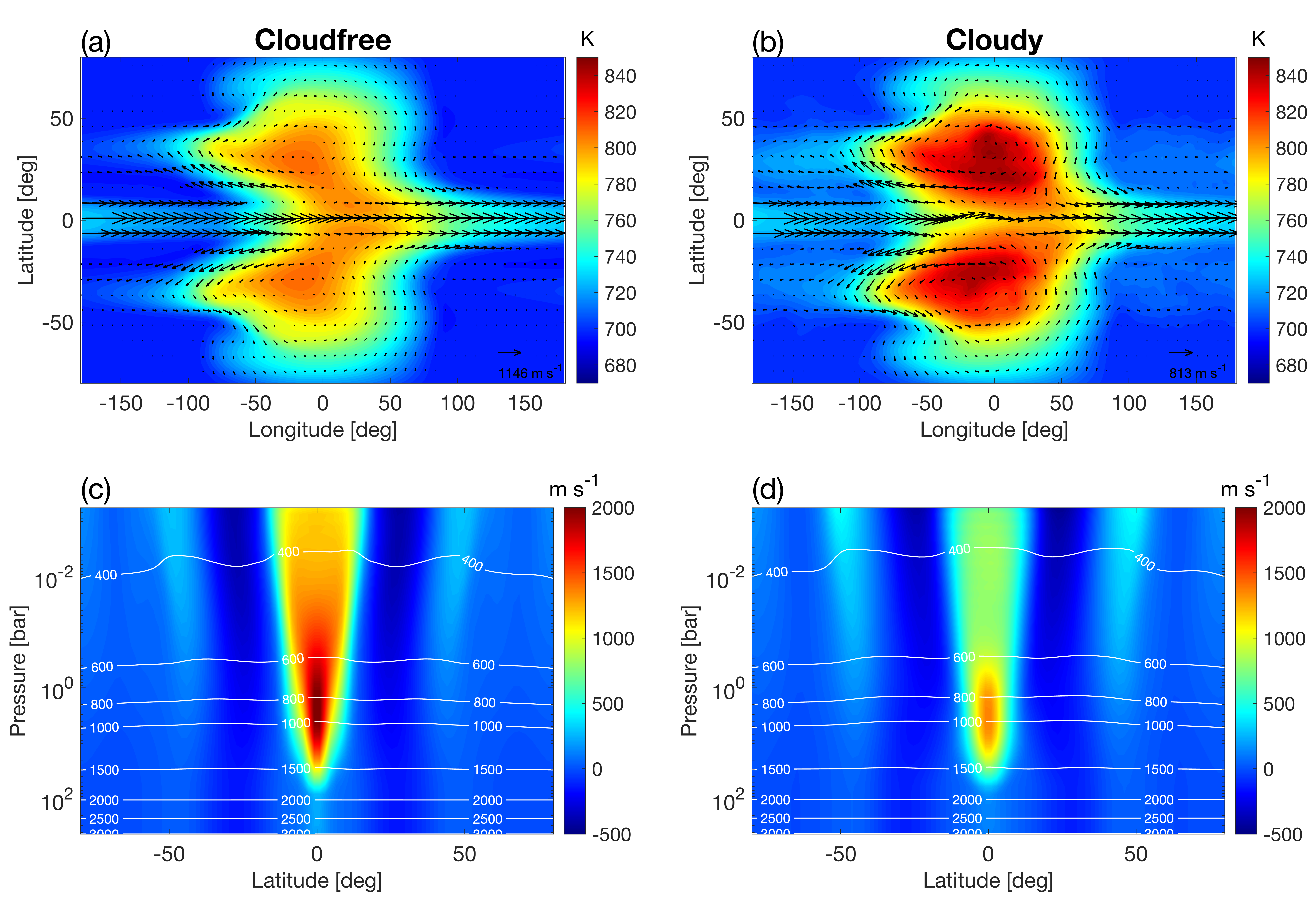}
    \caption{Atmospheric circulation under cloud-free (left) and cloudy (right) conditions in the ZTF0038B atmosphere. Top panels show horizontal temperature maps (colors, K) and horizontal winds (arrows, m\,s$^{-1}$) near the photosphere ($p\sim1$ bar). Bottom panels show zonal-mean zonal winds (colors, m\,s$^{-1}$) and temperatures (contours, K) as functions of latitude and pressure.}
    \label{fig:control-circ}
\end{figure*}

Measured atmospheric metallicities of field BDs are often close to solar values \citep[e.g.,][]{line2017}, and 1D atmospheric retrievals of ZTF0038B also favor metallicities consistent with solar \citep{broskilaing2026}. We therefore assume solar metallicity in our models, with a heat capacity of $c_p=13000$ J\,kg$^{-1}$\,K$^{-1}$ and a specific gas constant of $R=$ 3714 J\,kg$^{-1}$\,K$^{-1}$.
The horizontal resolution of the cubed sphere grid is C48, corresponding to 192$\times$96 grid points in longitude and latitude. The vertical domain spans pressures from 500 bar to $5\times10^{-4}$ bar using 53 levels evenly spaced in log pressure. At the lower boundary, temperatures are relaxed toward 3022 K to provide the adopted fiducial intrinsic temperature of $T_{\rm int}\approx920$ K.

All simulations are initialized from a state of rest and integrated using a dynamical time step of 10 s. The initial temperature--pressure profile transitions smoothly from a deep dry adiabat to an approximately isothermal upper atmosphere (Figure~\ref{fig:tp}). Statistical equilibrium is typically achieved after $\sim$1000 days, when the total kinetic energy and total energy fluctuate around statistically steady values without systematic long-term trends. Diagnostics are then computed from the final 100 days of model integration. Table \ref{tab:paras} summarizes the relevant parameters adopted in simulations.

\begin{figure*}
    \centering
    \includegraphics[width=0.95\textwidth]{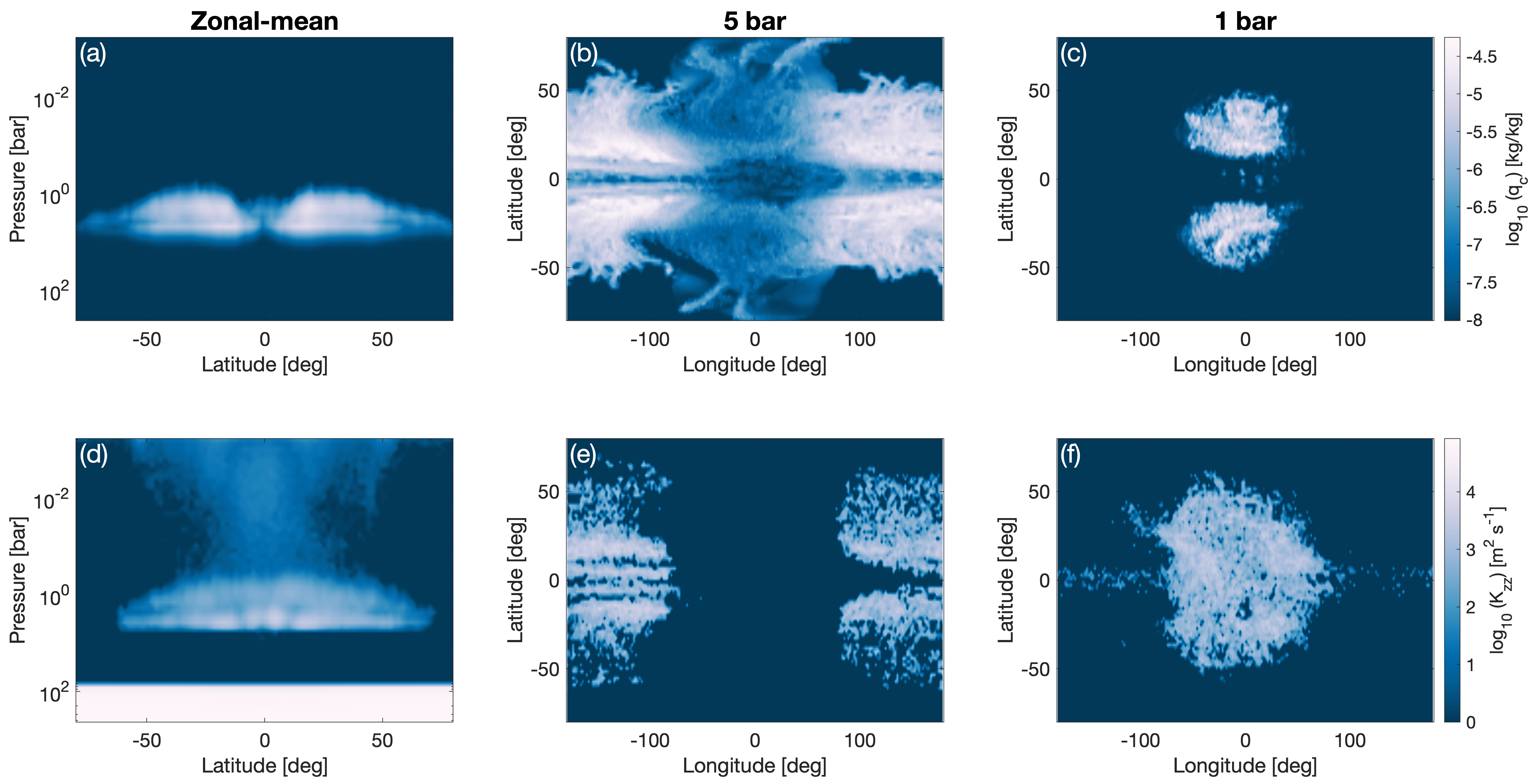}
    \caption{Cloud and vertical convective mixing ($K_{zz}$) structures in the cloudy ZTF0038B simulation. Top panels show the Na$_2$S cloud mass mixing ratio ($\log_{10} q_c$ [kg\,kg$^{-1}$]): (a) zonal-mean distribution, (b) horizontal distribution at $p\sim5$ bar, and (c) horizontal distribution at $p\sim1$ bar. Bottom panels show the corresponding vertical mixing coefficient, ($\log_{10} K_{zz}$ [m$^2$\,s$^{-1}$]: (d) zonal-mean distribution, (e) horizontal distribution at $p\sim5$ bar, and (f) horizontal distribution at $p\sim1$ bar.}
    \label{fig:control-cloud}
\end{figure*}

\begin{figure*}
    \centering
    \includegraphics[width=0.95\textwidth]{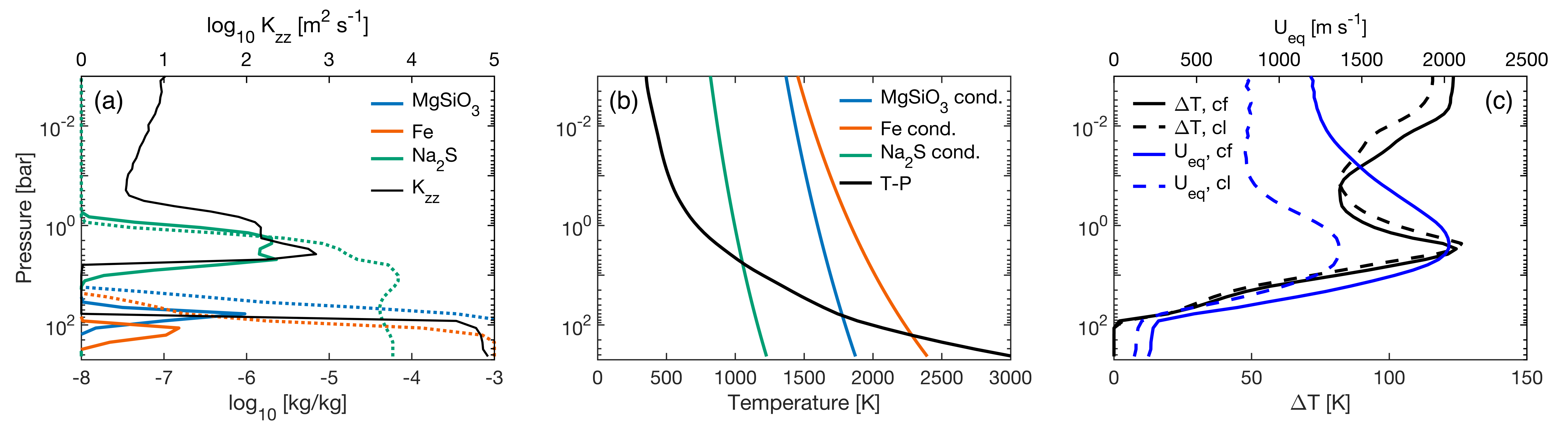}
    \caption{Panel (a): Global-mean vertical profiles of condensable vapor mixing ratios (colored dotted lines) and condensate mixing ratios (colored solid lines) for MgSiO$_3$, Fe, and Na$_2$S in the cloudy model. The black line shows the global-mean parameterized vertical mixing coefficient, $\log_{10}K_{zz}$ (m$^2$\,s$^{-1}$), referenced to the top axis. Panel (b): Global-mean temperature profile and condensation curves for MgSiO$_3$, Fe, and Na$_2$S in the cloudy model. Panel (c): Latitudinal-mean maximum day--night temperature contrast $\Delta T$ (K, black lines) and zonal-mean equatorial zonal wind $U_{eq}$ (m\,s$^{-1}$, blue lines) as functions of pressure. Solid and dashed curves represent the cloud-free and cloudy models, respectively. }
    \label{fig:control-clouddiag}
\end{figure*}

\begin{figure*}
    \centering
    \includegraphics[width=0.95\textwidth]{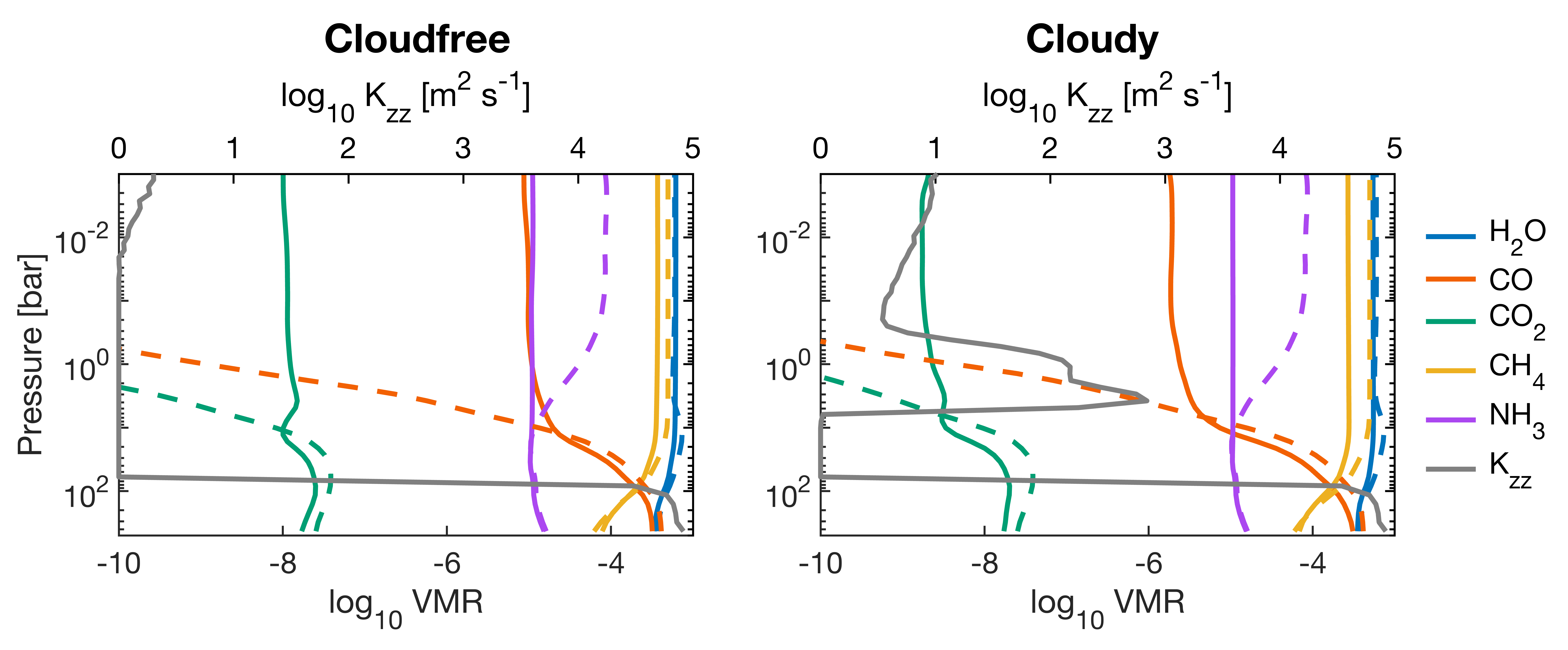}
    \caption{Global-mean vertical profiles of $\log_{10}K_{zz}$ (gray lines, m$^2$\,s$^{-1}$) and $\log_{10}$VMR for selected chemical species in the cloud-free (left) and cloudy (right) models. Solid curves show chemically disequilibrium abundances based on mini-chem calculations, and dashed curves show chemical-equilibrium abundances. }
    \label{fig:control-globalmean}
\end{figure*}

\begin{figure*}
    \centering
    \includegraphics[width=0.95\textwidth]{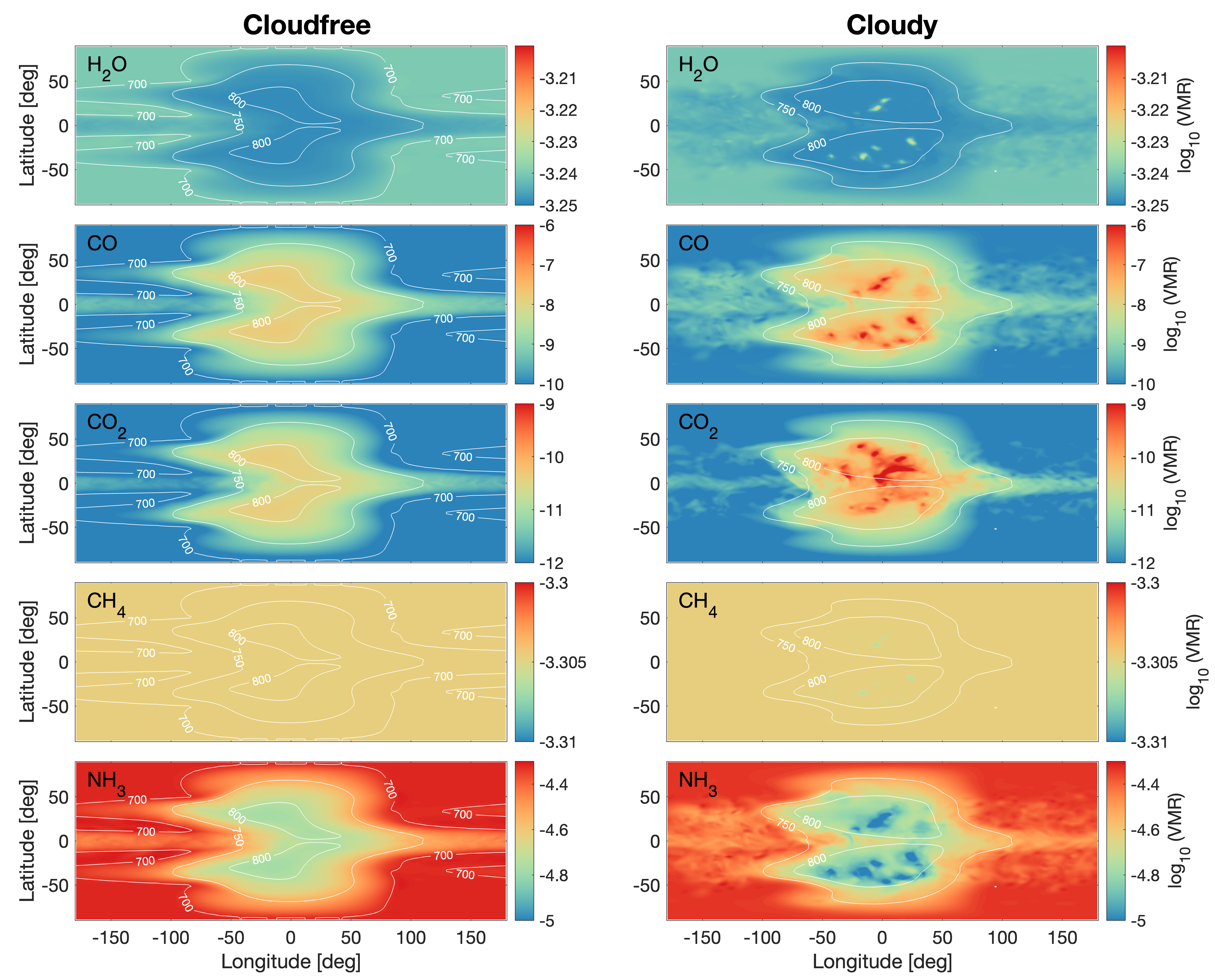}
    \caption{Horizontal distributions of H$_2$O, CO, CO$_2$, CH$_4$, and NH$_3$ under equilibrium chemistry (colors) at $p\sim1$ bar in the ZTF0038B atmosphere, overlaid with horizontal temperature maps at the same level with an interval of 50~K (contours). The left and right columns show the cloud-free and cloudy models, respectively. Contour values are labeled on the curves. }
    \label{fig:control-chemce}
\end{figure*}

\begin{figure*}
    \centering
    \includegraphics[width=0.95\textwidth]{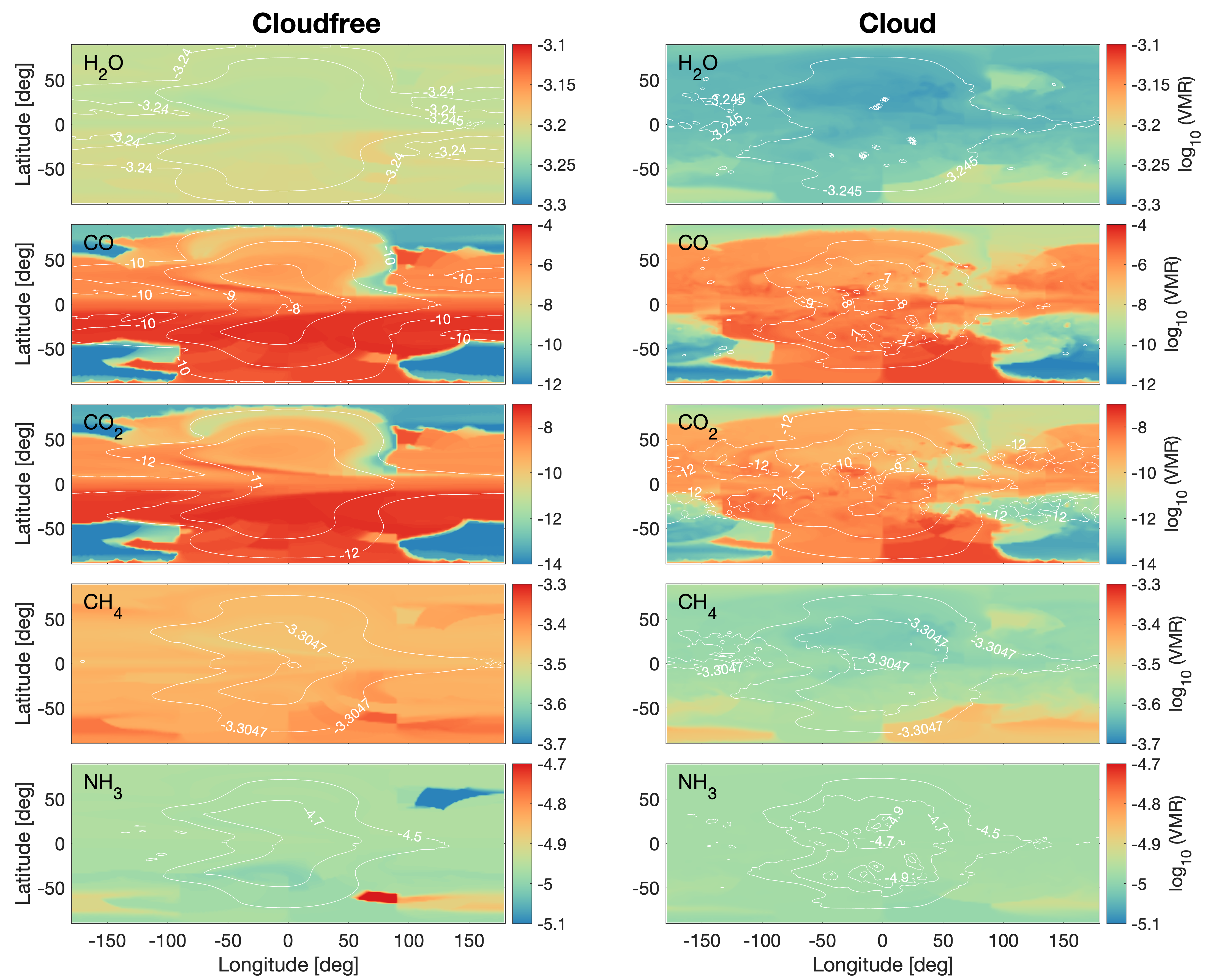}
    \caption{Horizontal distributions of H$_2$O, CO, CO$_2$, CH$_4$, and NH$_3$ at $p\sim1$ bar under disequilibrium chemistry (colors) and equilibrium chemistry (contours) in the ZTF0038B atmosphere. The left and right columns show the cloud-free and cloudy models, respectively. Contour values are labeled on the curves.}
    \label{fig:control-chem}
\end{figure*}

\section{Results}\label{sec:ztf_results} 

\subsection{Circulation and cloud structures}\label{sec:ztf_circ}

Figure~\ref{fig:control-circ} shows the atmospheric circulation pattern and temperature maps for the cloud-free and cloudy models. Regardless of cloud inclusion, the circulation is primarily driven by the strong day–night temperature contrast and exhibits a classical Matsuno–Gill-like pattern \citep{showman2011}, characterized by equatorial-trapped eastward-propagating Kelvin waves and westward-propagating Rossby waves. The excitation of planetary-scale equatorial waves transports angular momentum toward the equator, driving equatorial superrotation, where the angular momentum per unit mass exceeds that of the underlying planet. The equatorial eastward jet reaches $\sim$2000 m\,s$^{-1}$ in the cloud-free model and $\sim$1200 m\,s$^{-1}$ in the cloudy model, with horizontal temperature contrasts of less than 150~K near the photosphere ($p\sim1$ bar).

Because of the strong and inhomogeneous irradiation from the WD, the circulation of ZTF0038B more closely resembles that of irradiated HJs \citep[e.g.,][]{showman2002,pierrehumbert2019,showman2020review} rather than isolated BDs or directly imaged planets \citep[e.g.,][]{showman2013,tan2021a,tan2021b,tan2022bd}.
However, its much faster rotation relative to typical HJs compresses equatorial waves toward low latitudes, confining the Rossby waveguide within $\sim\pm$40$^\circ$ latitudes and producing a relatively narrow equatorial jet. This behavior is consistent with the fast-rotation regime identified by \citet{tan2020}, who systematically explored the transition from typical hot Jupiter rotation rates to those relevant for WD–BD systems. As a result, the temperature structure differs from that of HJs, with eastward hotspot offsets confined near the equator, while off-equatorial regions exhibit westward shifts associated with standing Rossby waves.

In the cloudy model, the overall cloud coverage is limited. The dominant cloud species, Na$_2$S, extends from $\sim$10 to 1 bar (Figure~\ref{fig:control-cloud}(a)), whereas MgSiO$_3$ and Fe clouds are largely absent from the photosphere and confined below $\sim$50 bar (Figure~\ref{fig:control-clouddiag}(a)). 
This depletion of MgSiO$_3$ and Fe clouds in the observable atmosphere arises primarily from efficient rainout at deeper levels due to high gravity, which removes most of their condensable vapor (Figure~\ref{fig:control-clouddiag}(a-b)). Vertical transport by large-scale circulation and convective mixing is insufficient to transport the condensates to lower pressures against sedimentation.
Such depletion of silicate and iron condensates in mid- and late-T dwarf atmospheres is consistent with previous observational and theoretical studies \citep{ackerman2001,apai2013,radigan2014}. 
By comparison, Na$_2$S condenses at shallower levels and remains abundant in the photosphere. Its horizontal distribution exhibits a strong pressure dependence, set by temperature variations relative to its condensation curve, with Na$_2$S concentrated on the nightside at $\sim$3–10 bar and shifting to the dayside at $\sim$0.5–3 bar (Figure~\ref{fig:control-cloud}(b-c)).

\begin{figure*}
    \centering
    \includegraphics[width=0.95\textwidth]{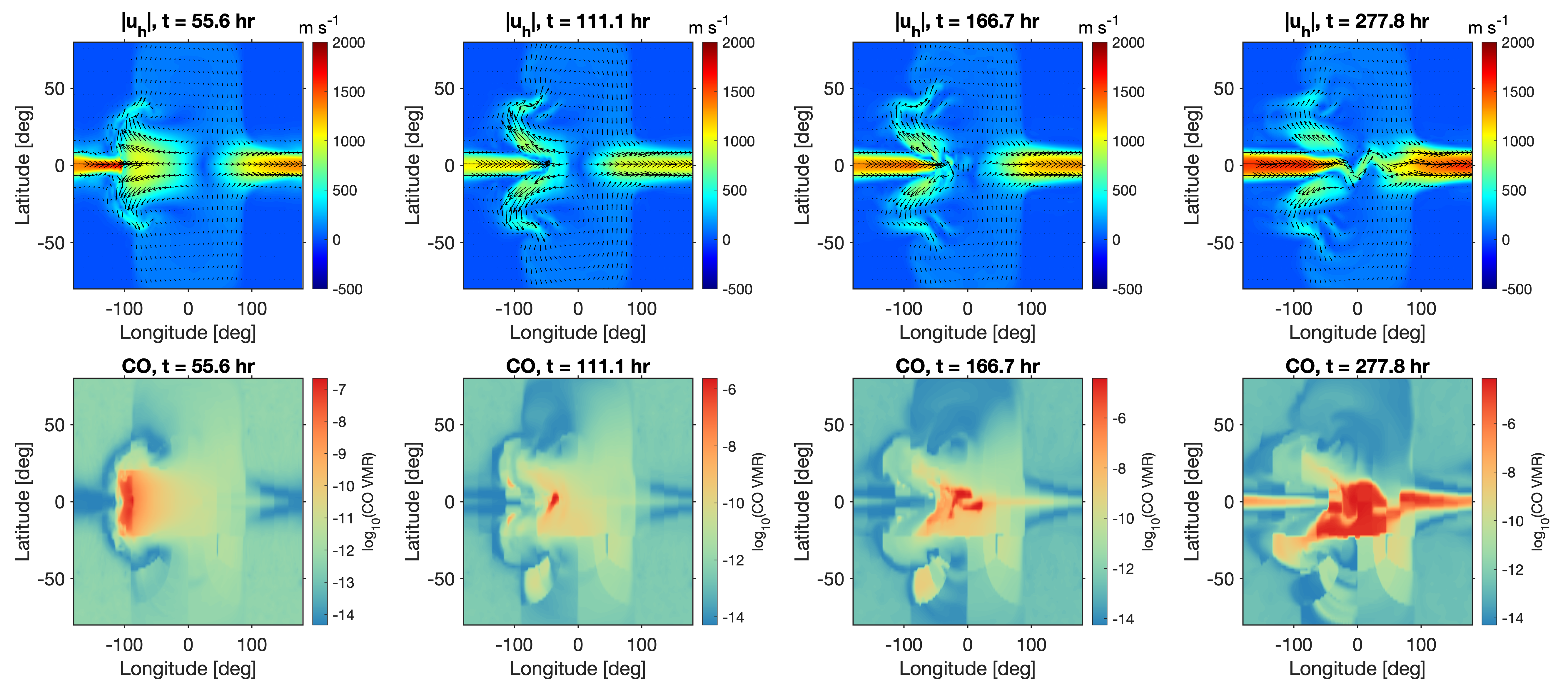}
    \caption{The breaking of north--south symmetry at $\sim$1 bar during the spin-up of the cloudy simulation. The top panels show the temporal evolution of the horizontal wind speed ($|\boldsymbol{u_h}|=\sqrt{u^2+v^2}$), with arrows indicating the horizontal flow direction. The bottom panels show the corresponding evolution of the CO abundance in $\log_{10}$ VMR. Similar hemispheric asymmetries develop in the other disequilibrium chemical species in the cloudy simulation and are also found in the cloud-free simulation.}
    \label{fig:asymmetry}
\end{figure*}

\subsection{Chemical structures}\label{sec:ztf_chem}
Figure~\ref{fig:control-globalmean} shows the global-mean VMRs of a few key gaseous chemical species, H$_2$O, CO, CO$_2$, CH$_4$, and NH$_3$, both from disequilibrium chemistry of mini-chem (solid lines) and from equilibrium chemistry (dashed lines), together with global-mean $K_{zz}$. The equilibrium profiles reflect the temperature dependence of the net chemical reactions \citep{tsai2021,tsai2022}: CO and CO$_2$ are favored at higher temperatures, whereas H$_2$O, CH$_4$, and NH$_3$ become more abundant at lower temperatures. Consequently, CO and CO$_2$ decrease with increasing pressure in equilibrium, whereas H$_2$O, CH$_4$, and NH$_3$ increase. 
The equilibrium abundances of CO and CO$_2$ are particularly sensitive to temperature in this regime, varying by up to ten orders of magnitude throughout the atmosphere. Such steep vertical abundance gradients imply that vertical transport can sample source regions with vastly different CO and CO$_2$ abundances, allowing spatial variations in upward and downward transport to generate large horizontal contrasts in the observable atmosphere. In contrast, the variation in H$_2$O, CH$_4$, and NH$_3$ are much weaker because their equilibrium abundances are less sensitive to temperature over this range.

Vertical transport strongly modifies these profiles under disequilibrium chemistry (solid lines in Figure~\ref{fig:control-globalmean}). The quench levels, below which the species remain close to chemical equilibrium, lie between $\sim10$ and $\sim100$~bar. This range is broadly consistent with the joint retrieval of the JWST spectra, which fits longitude-dependent quench pressures of $10^{1.2}$--$10^{1.4}$~bar \citep{broskilaing2026}. Above the quench levels, CO is transported upward from deeper layers, increasing its photospheric abundance from $\sim$10$^{-8}$ under chemical equilibrium to $\sim$10$^{-5}$ in the cloud-free disequilibrium model. CO$_2$ also increases in the observable atmosphere, from $\sim$10$^{-10}$ to $\sim$10$^{-8}$. In contrast, the abundances of H$_2$O, CH$_4$, and NH$_3$ are reduced relative to their equilibrium values by vertical mixing. In the cloudy model, the photospheric CO and CO$_2$ abundances are $\sim$10$^{-5.5}$ and $\sim$10$^{-8.5}$, respectively, lower than in the cloud-free case, indicating weaker upward transport in the presence of clouds.

The horizontal distributions further show how 3D transport reshapes the chemical structure in the observable atmosphere (Figures~\ref{fig:control-chemce} \& \ref{fig:control-chem}). Under chemical equilibrium, the large day--night temperature contrast in the ZTF0038B atmosphere produces pronounced chemical differences between the dayside and nightside. As shown in Figure~\ref{fig:control-chemce}, H$_2$O and NH$_3$ are depleted on the dayside, whereas CO and CO$_2$ are enhanced. Consistent with the steep equilibrium gradients seen in the global-mean profiles, the day--night contrasts in CO and CO$_2$ are much stronger than those of the other species because of their high temperature sensitivity.

Under disequilibrium chemistry, horizontal transport substantially reduces the day--night abundance contrasts, especially within the equatorial waveguide (Figure~\ref{fig:control-chem}). H$_2$O, CH$_4$, and NH$_3$ become nearly homogeneous globally, whereas CO and CO$_2$ retain order-of-magnitude spatial inhomogeneities. Within roughly $\pm50^{\circ}$ latitude, CO and CO$_2$ abundances exceed their local equilibrium values and are relatively homogeneous between the dayside and nightside, indicating upward transport from deeper carbon-rich layers followed by efficient horizontal redistribution. Beyond this latitude range, however, CO and CO$_2$ are strongly depleted and can even fall below their local equilibrium abundances, particularly on the nightside, likely due to downward transport of low-abundance air from higher altitudes.

In the cloud-free model, the local CO abundance within roughly $\pm50^{\circ}$ latitude reaches $\sim10^{-4}$, about four orders of magnitude above its equilibrium values there and nearly comparable to the disequilibrium CH$_4$ abundance. Such strong local enhancement may make disequilibrium opacity feedback particularly important in the CO absorption bands, and its impact on the broadband thermal structure deserves further investigation in future fully coupled radiative calculations. 

The meridional extent over which CO and CO$_2$ remain well mixed and abundant is closely tied to the large-scale circulation pattern. Horizontal mixing is efficient within the equatorial waveguide ($\pm50^{\circ}$ latitude), where the equatorial jet and associated Rossby wave activity dominate, but becomes much weaker at higher latitudes outside the waveguide. Consequently, CO- and CO$_2$-rich air is efficiently redistributed at low latitudes, whereas day--night exchange remains incomplete at high latitudes, allowing longitudinal abundance contrasts to persist. These inhomogeneities may leave observable imprints on phase curves in CO- and CO$_2$-sensitive bands, while bands dominated by H$_2$O, CH$_4$, and NH$_3$ are expected to remain comparatively uniform.

Such pronounced CO and CO$_2$ inhomogeneities in the ZTF0038B atmosphere differ from those in typical, more slowly rotating hot Jupiter atmospheres, where horizontal mixing tends to produce nearly uniform abundances globally \citep[e.g.,][]{mendoncca2018,drummond20183d,drummond2020,parmentier2026}. This behavior in ZTF0038B suggest the presence of strong mixing barriers associated with the rapidly rotating regime, which limits horizontal transport outside the equatorial waveguide. 

Latitudinal chemical variations can also arise in rapidly rotating isolated BDs through cloud-modified thermal structures. In such atmospheres, weak rotational constraints near the equator allow stronger vertical motions and preferential low-latitude cloud formation. The resulting equatorial cloud radiative feedback warms the low latitudes and generates an equator-to-pole thermal gradient \citep{vos2017,suarez2023,tan2021a,teinturier2026}. Under chemical equilibrium, this thermal structure can produce low-latitude depletion of CH$_4$, H$_2$O, and NH$_3$, together with CO and CO$_2$ enrichment \citep{charnay2025}. In Part II of this series (Lai \& Tan 2026, in prep.), we further show that these latitudinal chemical variations persist in rapidly rotating weakly irradiated, and even isolated, BDs when disequilibrium transport and horizontal quenching are included. Although this cloud-driven mechanism differs from the circulation-driven disequilibrium structures in ZTF0038B, both cases highlight the ability of rapid rotation to maintain observable latitudinal chemical inhomogeneities.

Such latitudinal variations may be detectable by comparing the apparent rotational broadening of different molecular lines in high-resolution spectra \citep{charnay2025}.  Complementary evidence comes from JWST time-resolved retrievals of the rapidly rotating T dwarf SIMP-0136, which inferred phase-dependent CO$_2$ and H$_2$S abundance variations correlated with atmospheric temperature variations \citep{nasedkin2025}. Together, these studies suggest that chemical inhomogeneities in rapidly rotating BDs can be probed through time-resolved and high-resolution spectroscopy.

Note that these gas species exhibit a clear north--south asymmetry in the observable atmosphere, reflecting an asymmetric circulation. During the spin-up, both the circulation and chemical distributions are initially approximately symmetric about the equator, but after $\sim$100 hr ($\sim$10 rotation periods), transient eddy activity breaks this symmetry and produces strongly non-zonal, hemispherically asymmetric flow near the substellar region (top panels of Figure~\ref{fig:asymmetry}). The resulting asymmetric meridional transport subsequently imprints a corresponding hemispheric asymmetry on the chemical distributions, with CO shown as a representative example in the bottom panels of Figure~\ref{fig:asymmetry}. This symmetry-breaking behavior occurs in both the cloud-free and cloudy simulations, suggesting that it is dynamically driven rather than cloud-induced. Similar spontaneous breakdowns of north--south symmetry under equatorially symmetric forcing have been reported in high-resolution hot Jupiter GCMs, where transient waves, vortices, and eddy activity contribute to asymmetric large-scale circulation \citep{menou2020,skinner2021}. In our rapidly rotating system, the strong equatorial jet and limited cross-equatorial transport further inhibit homogenization between the two hemispheres, allowing the chemically asymmetric state to persist.


\subsection{Cloud effects on circulation and mixing}\label{sec:cloud-effect}

The inclusion of clouds reduces the equatorial jet speed by weakening the day--night temperature contrast over the pressure range that most strongly drives the circulation ($\sim$3--100 bar; Figure~\ref{fig:control-clouddiag}(c)). This reduction is mainly caused by the dominant Na$_2$S clouds, which preferentially form on the nightside at $\sim$3--10 bar (Figure~\ref{fig:control-cloud}(b)). Their radiative feedback heats the atmosphere below the cloud layer, warms the nightside photosphere, and thereby reduces the thermal forcing of the large-scale flow. At deeper levels below $\sim$30 bar, MgSiO$_3$ clouds are more abundant on the dayside and reflect part of the incident shortwave flux, further reducing the dayside temperature contrast (not shown). In contrast, Fe clouds reside mostly below $\sim$100 bar, where stellar radiation is weak, and therefore have little radiative impact. Together, these cloud effects reduce the deep day--night thermal forcing, limit the initial acceleration of the flow, and lead to a lower peak jet speed in the cloudy model. Above $\sim$3 bar, the day--night temperature contrast becomes slightly larger than in the cloud-free model because of Na$_2$S-induced dayside warming (black curves in Figure~\ref{fig:control-clouddiag}(c)), but this occurs above the main region controlling the jet acceleration.

Cloud radiative effects also induce sub-grid convective mixing and lead to the formation of a secondary detached convective layer within the cloud-forming region (Figures~\ref{fig:control-cloud}(d--f) and \ref{fig:control-clouddiag}(a)). The dominant Na$_2$S cloud heats the atmosphere near the cloud base and cools it near the cloud top through radiative feedback, steepening the local temperature gradient and enhancing $K_{zz}$ within and above the cloud layer, roughly between 1 and 10 bar (Figure~\ref{fig:control-clouddiag}(a)). In turn, this detached convective layer provides strong vertical mixing that helps maintain the cloud layer against rapid gravitational settling and produces a sharp cloud top. Such coupling between cloud-radiative feedback and parameterized sub-grid mixing has also been reported in previous studies \citep{burrows2006,tan2019,lee2023}. Heating near the cloud base also stabilizes the adjacent layers, causing $K_{zz}$ to rapidly decrease toward its background value. At greater depths, $K_{zz}$ remains strong within the deep convective region below $\sim100$ bar.


Above $\sim$1 bar, where clouds are nearly absent, enhanced vertical mixing ($\log_{10}(K_{zz})>1$) still occurs in localized regions (Figure~\ref{fig:control-cloud}(d)). These regions are associated with local instability, indicated by relatively large lapse rates (i.e., $\partial \ln{T}/\partial \ln{p} \ge R/c_p$), and are often found near the morning terminator along the equator (not shown), where temperatures are locally minimized. The spatial distribution of vertical temperature gradient exhibits wave-like structures, suggesting that these instabilities are likely driven by large-scale wave activity rather than local convection. 

The parameterized $K_{zz}$ and resolved vertical velocity represent distinct transport processes: $K_{zz}$ describes unresolved local convective mixing within unstable regions, whereas the resolved vertical circulation directly advects chemical tracers on large scales. Clouds affect both pathways by enhancing $K_{zz}$ within the detached convective layer and modifying the large-scale circulation. In our simulations, the dominant Na$_2$S cloud layer lies mainly between $\sim$1 and 10 bar, largely above the CO and CO$_2$ quench level at $\sim10$ bar (Figure~\ref{fig:control-globalmean}(b)). The quench level is defined as the region where the transport timescale becomes shorter than the chemical conversion timescale and the abundances begin to depart from local chemical equilibrium. The disequilibrium abundances of CO and CO$_2$ are therefore controlled primarily by resolved vertical advection rather than by the cloud-induced $K_{zz}$. 
As the circulation weakens in the cloudy model (Figure~\ref{fig:control-circ}), large-scale vertical transport from the deep atmosphere is reduced. As a result, CO and CO$_2$ quench at shallower levels and have lower abundances above their quench levels than in the cloud-free model (Figure~\ref{fig:control-globalmean}).


Cloud radiative feedback also modifies the horizontal chemical distributions in the observable atmosphere. Under chemical equilibrium, the Na$_2$S-induced dayside warming near $p\sim1$ bar slightly enhances CO and CO$_2$ and depletes H$_2$O and CH$_4$ on the dayside relative to the cloud-free model. Under disequilibrium chemistry, however, the reduced large-scale transport lowers the photospheric CO and CO$_2$ abundances relative to the cloud-free case and weakens their horizontal homogenization. Consequently, CO and CO$_2$ remain more spatially inhomogeneous in the cloudy model.

\begin{figure*}
    \centering
    \includegraphics[width=0.94\linewidth]{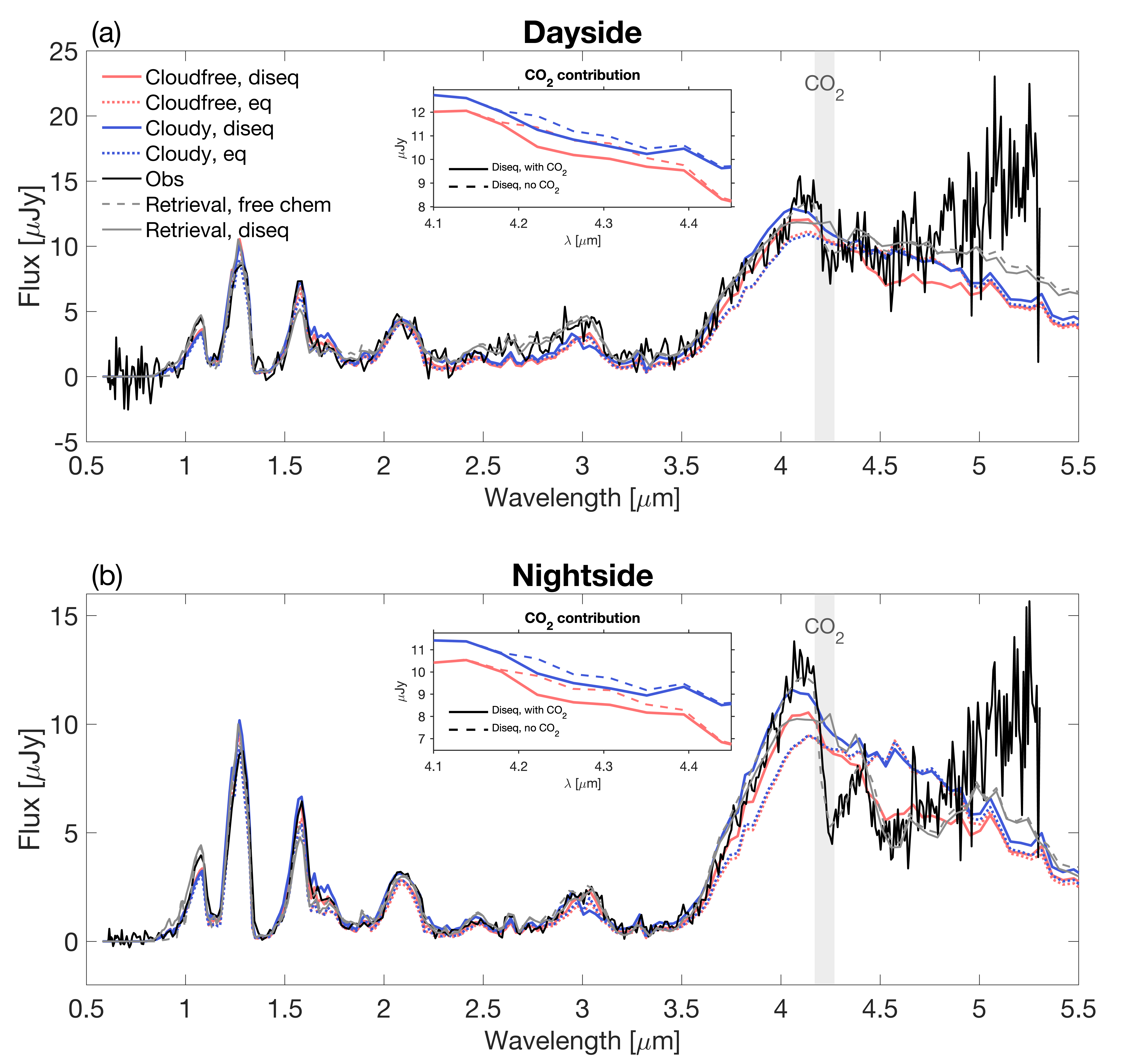}
    \caption{Emission flux as a function of wavelength for ZTF0038B on the dayside (a) and nightside (b). Red and blue curves show the GCM spectra from the cloud-free and cloudy models, respectively. For the GCM spectra, solid and dotted curves represent the spectra under disequilibrium and equilibrium chemistry, respectively. The black curve shows the observed spectrum. Gray curves show the best-fit retrieved spectra \citep{broskilaing2026}: the dashed gray curve is the free retrieval, in which molecular abundances are not constrained by chemistry, and the solid gray curve is the disequilibrium chemistry retrieval, in which chemical equilibrium is assumed with a free quench-pressure parameter. The insets zoom in on the 4.1--4.45 $\upmu$m region and compare the disequilibrium GCM spectra computed with (solid) and without CO$_2$ opacity (dashed). The shaded region marks the 4.17--4.27 $\upmu$m CO$_2$ bandpass used to construct the phase curves in Figure~\ref{fig:control-co2}.}
    \label{fig:emission}
\end{figure*}

\section{Observational signatures}\label{sec:ztf_spectra}
\begin{figure*}
    \centering
    \includegraphics[width=0.9\linewidth]{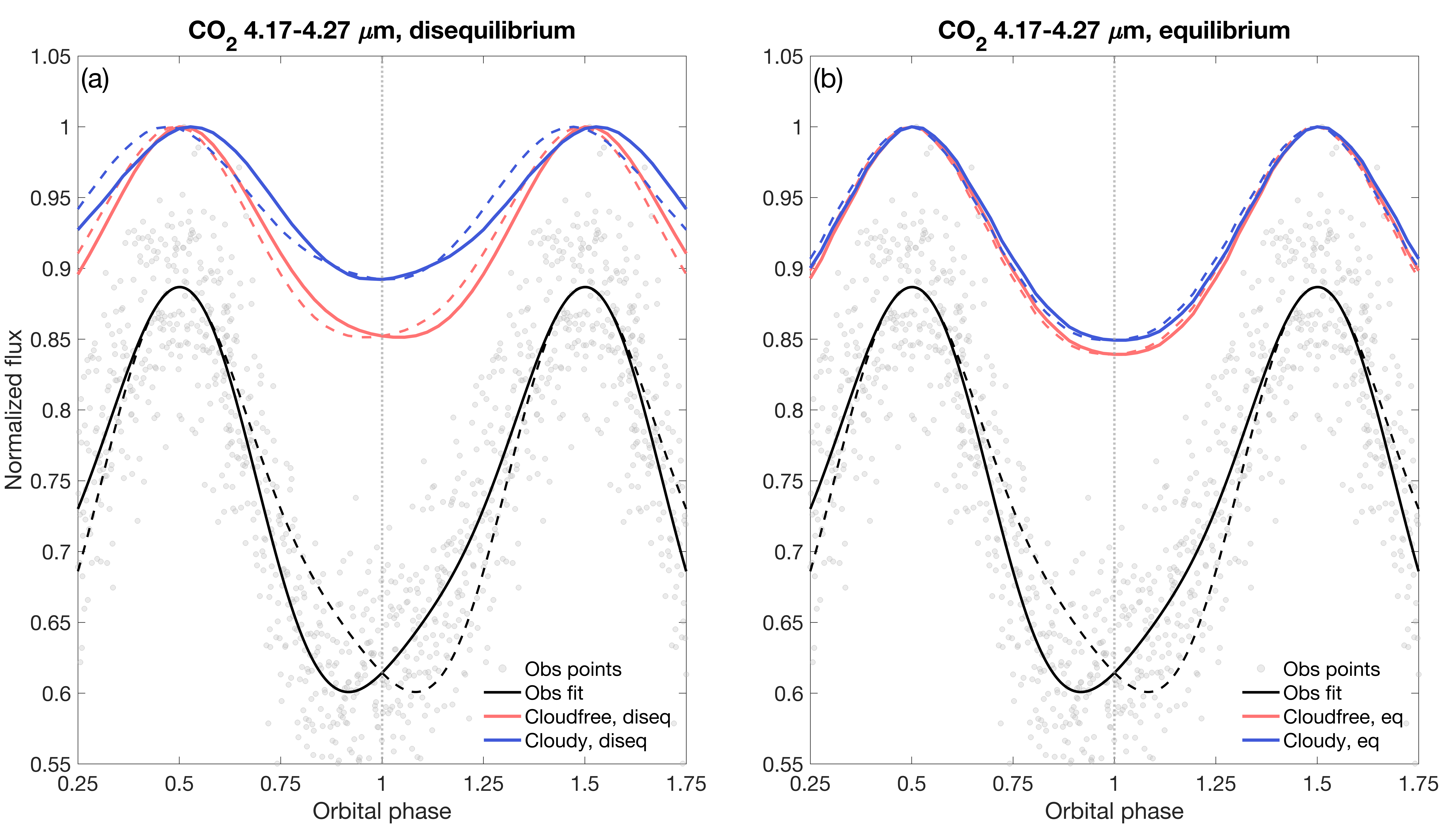}
    \caption{Normalized band-mean flux over 4.17--4.27 $\upmu$m, tracing the CO$_2$ absorption feature, as a function of orbital phase, for disequilibrium chemistry (a) and equilibrium chemistry (b). Red and blue curves denote the cloud-free and cloudy cases. Solid colored curves show the model phase curves, while dashed colored curves show the same curves reflected about phase = 0.5 to highlight the phase curve asymmetry. 
    Gray points show the JWST time-series band-mean fluxes from \citet{broskilaing2026}, normalized by the maximum observed band-mean flux. The black solid curve shows the weighted second-order Fourier fit to these normalized observations. The black dashed curve is not fitted independently, but is the same observational fit reflected about phase = 0.5, following the comparison used by \citet{broskilaing2026}, to illustrate departures from a phase curve symmetric about the substellar point.}
    \label{fig:control-co2}
\end{figure*}

We post-process the GCM outputs (temperature, cloud, and chemical abundances) using the radiative transfer model PICASO to compute synthetic observational signatures (Section~\ref{sec:picaso}), which are then compared with JWST time-resolved NIRSpec/PRISM spectra \citep{broskilaing2026}.
Figure~\ref{fig:emission} shows the thermal emission spectra on the dayside and nightside for both cloud-free and cloudy models. Figure~\ref{fig:control-co2} presents the normalized band-mean phase curve over 4.1--4.3~$\upmu$m, tracing the CO$_2$ absorption feature associated with the observed asymmetric phase variation.

\subsection{Emission spectra}\label{sec:emission}
Overall, the models reproduce the main spectral features between 0.5 and 4.5~$\upmu$m on both the dayside and nightside, including prominent near-infrared H$_2$O, CO, and CH$_4$ absorption bands (Figure~\ref{fig:emission}). 
However, the models do not fully reproduce the CO$_2$ feature centered near 4.2~$\upmu$m, whose observed depth requires substantially higher CO$_2$ abundances than predicted by our disequilibrium-chemistry calculations. 
This is broadly consistent with the spectroscopic retrieval results of \citet{broskilaing2026}. Their disequilibrium-chemistry retrievals (solid gray curves in Figure~\ref{fig:emission}) provide spectral fits similar to our GCM spectra overall, but both show difficulty matching the observed CO$_2$ feature, which is better reproduced by the free-chemistry retrieval (dashed gray curves in Figure~\ref{fig:emission}).
Because the free retrieval does not chemically constrain molecular abundances, it infers a CO$_2$ abundance about three orders of magnitude higher than the disequilibrium-chemistry retrieval. 

The origin of this enhanced CO$_2$ abundance remains unclear. The near-solar metallicity and C/O ratio inferred by \citet{broskilaing2026} suggest that an unusual bulk composition is unlikely to be the primary explanation. Our results further indicate that dynamical transport and quenching alone do not produce sufficient CO$_2$ to match the observed feature, pointing to additional chemical pathways or other missing physical processes that are not captured in the present model.
Beyond the CO$_2$ feature, the increase in thermal flux toward wavelengths longer than $\sim$4.5~$\upmu$m is also not captured by either the models or the retrievals.


Disequilibrium and equilibrium spectra show only minor differences across most wavelength bands (red and blue curves in Figure~\ref{fig:emission}). This is because the dominant infrared absorbers, including H$_2$O and CH$_4$, have similar abundances in the observable atmosphere in both cases (Figure~\ref{fig:control-globalmean}). The main spectral differences are therefore confined to specific CO and CO$_2$ absorption bands.
The enhanced photospheric CO abundance produced by disequilibrium chemistry strengthens the absorption feature near 4.4~$\upmu$m, improving the agreement between the cloud-free disequilibrium spectrum and the observed nightside CO feature (Figure~\ref{fig:emission}(b)). However, the CO$_2$ feature near 4.2~$\upmu$m remains underpredicted despite the enhanced CO$_2$ abundance.

Because of the limited cloud coverage in the observable atmosphere, the cloudy spectra remain broadly similar to the cloud-free case, particularly for the equilibrium-chemistry spectra, which are mainly controlled by the temperature structure. The most prominent difference appears near 4.4--4.6~$\upmu$m in the disequilibrium-chemistry spectra: the cloud-free model shows stronger CO absorption owing to stronger vertical transport and a higher upper-atmosphere CO abundance (Figure~\ref{fig:control-globalmean}), which better matches the observed CO feature, whereas weaker transport in the cloudy model reduces this feature.

\subsection{Asymmetric CO$_2$ phase curve}\label{sec:phasecurve}

JWST/NIRSpec PRISM phase-resolved spectroscopy of ZTF0038B revealed a pronounced wavelength-dependent phase curve asymmetry associated with the CO$_2$ absorption band near 4.2~$\upmu$m (gray points and black curves in Figure~\ref{fig:control-co2}; \citealt{broskilaing2026}). Unlike the nearly symmetric broadband thermal phase curve, the band-integrated flux across this CO$_2$ feature shows a strong nightside asymmetry, with deeper absorption on the western nightside and weaker absorption on the eastern nightside. 
The retrieval analysis favored nearly longitudinally homogeneous abundances for CO$_2$ and other detected species, suggesting that the observed asymmetry may not require large-scale abundance variations within the retrieval framework.

Motivated by this observational feature, we compute band-mean phase curves in the CO$_2$ absorption band from our cloud-free and cloudy GCM spectra under both equilibrium and disequilibrium chemistry (Figure~\ref{fig:control-co2}). 
As shown in the inset of Figure~\ref{fig:emission}, the CO$_2$ absorption feature spans approximately 4.17--4.4~$\upmu$m. For direct comparison with the JWST CO$_2$ phase curve of \citet{broskilaing2026}, we adopt their 4.17--4.27~$\upmu$m bandpass. A broader 4.17--4.4~$\upmu$m bandpass yields a similar phase curve asymmetry. At each orbital phase, PICASO computes the emergent thermal intensity over the visible hemisphere from the 3D temperature, gas, and cloud-opacity fields. The resulting disk-integrated flux is then averaged over 4.17--4.27~$\upmu$m and normalized by its maximum value over the orbit to obtain the relative flux.
This comparison tests whether the simulated 3D thermal, cloud, and chemical structures can reproduce the observed CO$_2$ phase curve asymmetry.

Overall, the modeled CO$_2$-band phase curves have modest amplitudes in both the cloud-free and cloudy cases, with normalized flux variations smaller than $\sim$0.15, substantially below the observed phase amplitude of $\sim$0.3 (Figure~\ref{fig:control-co2}). The weaker modeled amplitude likely arises from the lower CO$_2$ abundance predicted by the models relative to the observations, which shifts the CO$_2$ photosphere to deeper layers with smaller horizontal temperature contrasts. Note that the observational fitting curve peaks below unity because the JWST data are normalized before fitting, whereas the fitted curve itself is not independently renormalized, producing an apparent vertical offset from the normalized GCM phase curves.

The disequilibrium-chemistry phase curves show a weak nightside asymmetry relative to the JWST observations. This asymmetry is absent under chemical equilibrium, indicating that it is not produced by the thermal structure or cloud opacity alone, but instead arises from spatially inhomogeneous CO$_2$ opacity associated with disequilibrium transport. In ZTF0038B, rapid rotation limits horizontal homogenization and allows circulation-driven vertical transport to maintain spatial CO$_2$ inhomogeneities in the photosphere (Figure~\ref{fig:control-chem}).
This circulation-driven chemical inhomogeneity provides a physical explanation for how spatial CO$_2$ variations can arise, which may not be captured by the longitudinally homogeneous CO$_2$ distribution favored by the retrieval analysis of \citet{broskilaing2026}.

The phase of the asymmetry differs between the two disequilibrium models: in the cloud-free case, the flux minimum occurs on the eastern nightside, whereas in the cloudy case it shifts to the western nightside, closer to the JWST phase curve. In the cloudy model, this asymmetry is weaker than in the cloud-free case, consistent with its reduced CO$_2$ abundance. A similar modeled nightside asymmetry also appears in the 4.4--4.6~$\upmu$m CO band (not shown), although no corresponding asymmetry is evident in the JWST data \citep{broskilaing2026}.

The closer agreement of the cloudy disequilibrium model with the observed CO$_2$ phase curve suggests that both cloud radiative feedback and transport-induced disequilibrium chemistry contribute to setting the asymmetry, although the modeled amplitude remains too small. A larger and more spatially inhomogeneous CO$_2$ abundance would strengthen the phase curve asymmetry. In our simulations, however, cloud-driven convective mixing occurs mainly above the CO$_2$ quench level and therefore has limited influence on the quenched photospheric abundance. If clouds were to form deeper and overlap the CO$_2$ quench region, stronger cloud-induced $K_{zz}$ could enhance upward transport and increase the photospheric CO$_2$ abundance. More efficient large-scale vertical transport could have a similar effect. In addition, missing chemical pathways and the neglected radiative feedback of disequilibrium chemistry could modify the CO$_2$ abundance, horizontal temperature structure, circulation, and cloud distribution. These possibilities warrant further investigation as potential pathways toward explaining the observed CO$_2$ phase curve asymmetry.

\section{Conclusion}\label{sec:conclude}

Using an updated 3D framework, \texttt{ADAM}, which couples the MITgcm dynamical core with the radiative transfer model SPARC, a miniature kinetic chemistry scheme (mini-chem), and a tracer-based cloud module, we investigate the atmosphere of ZTF0038B, an irradiated BD in a compact WD-BD system. We perform both cloud-free and cloudy simulations including MgSiO$_3$, Fe, and Na$_2$S clouds, and compute observable spectra and phase curves using PICASO. Our main findings are summarized as follows.

\begin{enumerate}
\item Under comparable stellar and internal heat fluxes, ZTF0038B develops a standing Matsuno--Gill wave response and an equatorial superrotating jet driven by day--night thermal forcing. Rapid rotation confines the circulation meridionally relative to canonical hot Jupiters, producing more symmetric broadband light curves without a significant hotspot offset (Section~\ref{sec:ztf_circ}). 

\item Circulation-driven disequilibrium chemistry enhances global-mean photospheric CO and CO$_2$ while slightly depleting H$_2$O, CH$_4$, and NH$_3$ relative to chemical equilibrium. Although most species are globally homogenized, rapid rotation limits horizontal homogenization and allows circulation-driven vertical transport to maintain strong horizontal CO and CO$_2$ inhomogeneities in the observable atmosphere (Section~\ref{sec:ztf_chem}). These chemical inhomogeneities produce asymmetric phase curves in CO and CO$_2$ absorption bands that could qualitatively explain the observed CO$_2$ asymmetry (Section~\ref{sec:phasecurve}). This provides evidence for an additional vertical mixing pathway driven by large-scale circulation, complementing the horizontal advection commonly invoked for HJs and the parameterized vertical diffusive mixing adopted in many BDs and 1D models. The disequilibrium spectra reproduce the main JWST molecular absorption features but underpredict the CO$_2$ band near 4.2~$\upmu$m (Section~\ref{sec:emission}).

\item In cloudy simulations, Na$_2$S clouds dominate the observable photosphere, whereas MgSiO$_3$ and Fe clouds rain out at deeper levels under the high gravity of ZTF0038B (Section~\ref{sec:ztf_circ}). Cloud radiative feedback reduces the day--night temperature contrast over the main circulation-driving region, weakens the large-scale circulation, and enhances localized convective mixing. However, because the dominant Na$_2$S clouds reside mostly above the CO and CO$_2$ quench region, cloud-driven convection has limited influence on the quenched abundances. Instead, the quenched CO and CO$_2$ abundances are controlled primarily by large-scale circulation. The weakened circulation in the cloudy model reduces the photospheric CO and CO$_2$ abundances and modifies the phase and amplitude of the CO$_2$-band light-curve asymmetry (Sections~\ref{sec:cloud-effect} and \ref{sec:ztf_spectra}).

\end{enumerate}

In Part I of this program, we presented dynamically coupled 3D cloud--chemistry simulations for a rapidly rotating irradiated atmosphere, using ZTF0038B as a representative case study. In Part II, we extend this framework to explore how the coupling among irradiation, clouds, dynamics, and chemistry varies across a broader range of irradiation levels, from isolated to strongly irradiated BDs (Lai \& Tan 2026, in prep.). These calculations investigate the transition between externally forced and internally driven atmospheric regimes and the resulting impact on chemical transport.

\section*{Acknowledgments}
XT is supported by the National Natural Science Foundation of China (grant No. 42475131).  YL is supported by the National Natural Science Foundation of China (grant No. 42508019).
This project is supported in part by Office of Science and Technology, Shanghai Municipal Government (grant Nos. 24DX1400100, ZJ2023-ZD-001). We acknowledge the computational support provided by the Siyuan-1 cluster, supported by the Center for High Performance Computing at Shanghai Jiao Tong University. JRF is supported by the ERC/UKRI Frontier Research Guarantee programme (EP/Z000327/1/CandY).
The JWST/NIRSpec observations used for comparison in this work are publicly available through MAST under \url{doi:10.17909/0z03-1m62}.

\bibliography{main}{}

@article{madhusudhan2018,
  title={Atmospheric retrieval of exoplanets},
  author={Madhusudhan, Nikku},
  journal={arXiv preprint arXiv:1808.04824},
  doi={https://doi.org/10.48550/arXiv.1808.04824},
  year={2018}
}

@article{barstow2020,
  title={Outstanding challenges of exoplanet atmospheric retrievals},
  author={Barstow, Joanna K and Heng, Kevin},
  journal={Space science reviews},
  volume={216},
  number={5},
  pages={82},
  year={2020},
  url={https://doi.org/10.1007/s11214-020-00666-x},
  publisher={Springer}
}

@article{van2021,
  title={ZTFJ0038+ 2030: A long-period eclipsing white dwarf and a substellar companion},
  author={van Roestel, Jan and Kupfer, Thomas and Bell, Keaton J and Burdge, Kevin and Mr{\'o}z, Przemek and Prince, Thomas A and Bellm, Eric C and Drake, Andrew and Dekany, Richard and Mahabal, Ashish A and others},
  journal={The Astrophysical Journal Letters},
  volume={919},
  number={2},
  pages={L26},
  year={2021},
  DOI={10.3847/2041-8213/ac22b7},
  publisher={IOP Publishing}
}

@article{beuermann2013,
  title={The eclipsing post-common envelope binary CSS21055: a white dwarf with a probable brown-dwarf companion},
  author={Beuermann, Klaus and Dreizler, Stefan and Hessman, Frederic V and Backhaus, U and Boesch, A and Husser, T-O and Nortmann, L and Schmelev, A and Springer, R},
  journal={Astronomy \& Astrophysics},
  volume={558},
  pages={A96},
  year={2013},
  doi={https://doi.org/10.1051/0004-6361/201322241},
  publisher={EDP Sciences}
}

@article{farihi2017,
  title={A circumbinary debris disk in a polluted white dwarf system},
  author={Farihi, J and Parsons, SG and G{\"a}nsicke, BT},
  journal={Nature Astronomy},
  volume={1},
  number={3},
  pages={0032},
  year={2017},
  doi={https://doi.org/10.1038/s41550-016-0032},
  publisher={Nature Publishing Group UK London}
}

@article{parsons2017,
  title={Two white dwarfs in ultrashort binaries with detached, eclipsing, likely sub-stellar companions detected by K2},
  author={Parsons, SG and Hermes, JJ and Marsh, TR and G{\"a}nsicke, Boris T and Tremblay, P-E and Littlefair, SP and Sahman, DI and Ashley, RP and Green, M and Rattanasoon, S and others},
  journal={Monthly Notices of the Royal Astronomical Society},
  volume={471},
  number={1},
  pages={976--986},
  year={2017},
  url={https://doi.org/10.1093/mnras/stx1610},
  publisher={Oxford University Press}
}

@article{casewell2018,
  title={The first sub-70 min non-interacting WD--BD system: EPIC212235321},
  author={Casewell, SL and Braker, IP and Parsons, SG and Hermes, JJ and Burleigh, MR and Belardi, C and Chaushev, A and Finch, NL and Roy, M and Littlefair, SP and others},
  journal={Monthly Notices of the Royal Astronomical Society},
  volume={476},
  number={1},
  pages={1405--1411},
  year={2018},
  doi={https://doi.org/10.1093/mnras/sty245},
  publisher={Oxford University Press}
}

@article{saumon2000,
  title={Molecular Abundances in the Atmosphere of the T Dwarf GL 229B},
  author={Saumon, D and Geballe, TR and Leggett, SK and Marley, MS and Freedman, RS and Lodders, K and Fegley, Jr, B and Sengupta, SK},
  journal={The Astrophysical Journal},
  volume={541},
  number={1},
  pages={374--389},
  DOI={10.1086/309410},
  year={2000}
}

@ARTICLE{line2017,
       author = {{Line}, Michael R. and {Marley}, Mark S. and {Liu}, Michael C. and {Burningham}, Ben and {Morley}, Caroline V. and {Hinkel}, Natalie R. and {Teske}, Johanna and {Fortney}, Jonathan J. and {Freedman}, Richard and {Lupu}, Roxana},
        title = "{Uniform Atmospheric Retrieval Analysis of Ultracool Dwarfs. II. Properties of 11 T dwarfs}",
      journal = {\apj},
         year = 2017,
        month = oct,
       volume = {848},
       number = {2},
          eid = {83},
        pages = {83},
          doi = {10.3847/1538-4357/aa7ff0},
archivePrefix = {arXiv},
       eprint = {1612.02809},
 primaryClass = {astro-ph.SR},
       adsurl = {https://ui.adsabs.harvard.edu/abs/2017ApJ...848...83L}
}

@article{miles2023,
  title={The JWST Early-release Science Program for Direct Observations of Exoplanetary Systems II: A 1 to 20 $\mu$ m Spectrum of the Planetary-mass Companion VHS 1256--1257 b},
  author={Miles, Brittany E and Biller, Beth A and Patapis, Polychronis and Worthen, Kadin and Rickman, Emily and Hoch, Kielan KW and Skemer, Andrew and Perrin, Marshall D and Whiteford, Niall and Chen, Christine H and others},
  journal={The Astrophysical journal letters},
  volume={946},
  number={1},
  pages={L6},
  year={2023},
  DOI={10.3847/2041-8213/acb04a},
  publisher={The American Astronomical Society}
}

@article{miles2020,
  title={Observations of disequilibrium CO chemistry in the coldest brown dwarfs},
  author={Miles, Brittany E and Skemer, Andrew JI and Morley, Caroline V and Marley, Mark S and Fortney, Jonathan J and Allers, Katelyn N and Faherty, Jacqueline K and Geballe, Thomas R and Visscher, Channon and Schneider, Adam C and others},
  journal={The Astronomical Journal},
  volume={160},
  number={2},
  pages={63},
  year={2020},
  DOI={10.3847/1538-3881/ab9114},
  publisher={The American Astronomical Society}
}

@article{sorahana2012,
  title={AKARI Observations of Brown Dwarfs. III. CO, CO2, and CH4 Fundamental Bands and Physical Parameters},
  author={Sorahana, S and Yamamura, I},
  journal={The Astrophysical Journal},
  volume={760},
  number={2},
  pages={151},
  year={2012},
  DOI={10.1088/0004-637X/760/2/151},
  publisher={The American Astronomical Society}
}

@article{cushing2011,
  title={The discovery of Y dwarfs using data from the wide-field infrared survey explorer (WISE)},
  author={Cushing, Michael C and Kirkpatrick, J Davy and Gelino, Christopher R and Griffith, Roger L and Skrutskie, Michael F and Mainzer, A and Marsh, Kenneth A and Beichman, Charles A and Burgasser, Adam J and Prato, Lisa A and others},
  journal={The Astrophysical Journal},
  volume={743},
  number={1},
  pages={50},
  year={2011},
  DOI={10.1088/0004-637X/743/1/50},
  publisher={The American Astronomical Society}
}

@article{metchev2015,
  title={Weather on other worlds. II. Survey results: spots are ubiquitous on L and T dwarfs},
  author={Metchev, Stanimir A and Heinze, Aren and Apai, D{\'a}niel and Flateau, Davin and Radigan, Jacqueline and Burgasser, Adam and Marley, Mark S and Artigau, {\'E}tienne and Plavchan, Peter and Goldman, Bertrand},
  journal={The Astrophysical Journal},
  volume={799},
  number={2},
  pages={154},
  year={2015},
  DOI={10.1088/0004-637X/799/2/154},
  publisher={The American Astronomical Society}
}

@article{artigau2009,
  title={Photometric variability of the T2. 5 brown dwarf SIMP J013656. 5+ 093347: Evidence for evolving weather patterns},
  author={Artigau, {\'E}tienne and Bouchard, Sandie and Doyon, Ren{\'e} and Lafreni{\`e}re, David},
  journal={The Astrophysical Journal},
  volume={701},
  number={2},
  pages={1534--1539},
  year={2009},
  DOI={10.1088/0004-637X/701/2/1534},
  publisher={The American Astronomical Society}
}

@article{burrows2006,
  title={L and T dwarf models and the L to T transition},
  author={Burrows, Adam and Sudarsky, David and Hubeny, Ivan},
  journal={The Astrophysical Journal},
  volume={640},
  number={2},
  pages={1063--1077},
  DOI={10.1086/500293},
  year={2006}
}

@article{ackerman2001,
  title={Precipitating condensation clouds in substellar atmospheres},
  author={Ackerman, Andrew S and Marley, Mark S},
  journal={The Astrophysical Journal},
  volume={556},
  number={2},
  pages={872--884},
  DOI={10.1086/321540},
  year={2001}
}

@ARTICLE{Teinturier2024,
       author = {{Teinturier}, L. and {Charnay}, B. and {Spiga}, A. and {B{\'e}zard}, B. and {Leconte}, J. and {Mechineau}, A. and {Ducrot}, E. and {Millour}, E. and {Cl{\'e}ment}, N.},
        title = "{The radiative and dynamical impact of clouds in the atmosphere of the hot Jupiter WASP-43 b}",
      journal = {\aap},
         year = 2024,
        month = mar,
       volume = {683},
          eid = {A231},
        pages = {A231},
          doi = {10.1051/0004-6361/202347069},
archivePrefix = {arXiv},
       eprint = {2401.14083},
 primaryClass = {astro-ph.EP},
       adsurl = {https://ui.adsabs.harvard.edu/abs/2024A&A...683A.231T}
}

@ARTICLE{Kennedy2025,
       author = {{Kennedy}, Thomas D. and {Rauscher}, Emily and {Malsky}, Isaac and {Roman}, Michael T. and {Beltz}, Hayley},
        title = "{Radiatively Active Clouds and Magnetic Effects Explored in a Grid of Hot Jupiter GCMs}",
      journal = {\apj},
         year = 2025,
        month = jan,
       volume = {978},
       number = {1},
          eid = {82},
        pages = {82},
          doi = {10.3847/1538-4357/ad9394},
archivePrefix = {arXiv},
       eprint = {2410.23436},
 primaryClass = {astro-ph.EP},
       adsurl = {https://ui.adsabs.harvard.edu/abs/2025ApJ...978...82K}
}

@article{bordwell2018,
  title={Convective dynamics and disequilibrium chemistry in the atmospheres of giant planets and brown dwarfs},
  author={Bordwell, Baylee and Brown, Benjamin P and Oishi, Jeffrey S},
  journal={The Astrophysical Journal},
  volume={854},
  number={1},
  pages={8},
  year={2018},
  DOI={10.3847/1538-4357/aaa551},
  publisher={The American Astronomical Society}
}

@article{leggett2021,
  title={Measuring and Replicating the 1--20 $\mu$ m Energy Distributions of the Coldest Brown Dwarfs: Rotating, Turbulent, and Nonadiabatic Atmospheres},
  author={Leggett, SK and Tremblin, Pascal and Phillips, Mark W and Dupuy, Trent J and Marley, Mark and Morley, Caroline and Schneider, Adam and Caselden, Dan and Guillaume, Colin and Logsdon, Sarah E},
  journal={The Astrophysical Journal},
  volume={918},
  number={1},
  pages={11},
  year={2021},
  DOI={10.3847/1538-4357/ac0cfe},
  publisher={The American Astronomical Society}
}

@article{steinrueck2019,
  title={The effect of 3D transport-induced disequilibrium carbon chemistry on the atmospheric structure, phase curves, and emission spectra of hot Jupiter HD 189733b},
  author={Steinrueck, Maria E and Parmentier, Vivien and Showman, Adam P and Lothringer, Joshua D and Lupu, Roxana E},
  journal={The Astrophysical Journal},
  volume={880},
  number={1},
  pages={14},
  year={2019},
  DOI={10.3847/1538-4357/ab2598},
  publisher={The American Astronomical Society}
}

@article{drummond20183d,
  title={The 3d thermal, dynamical, and chemical structure of the atmosphere of HD 189733b: Implications of wind-driven chemistry for the emission phase curve},
  author={Drummond, Benjamin and Mayne, Nathan J and Manners, James and Baraffe, Isabelle and Goyal, Jayesh and Tremblin, Pascal and Sing, David K and Kohary, Krisztian},
  journal={The Astrophysical Journal},
  volume={869},
  number={1},
  pages={28},
  year={2018a},
  DOI={10.3847/1538-4357/aaeb28},
  publisher={The American Astronomical Society}
}

@article{drummond2020,
  title={Implications of three-dimensional chemical transport in hot Jupiter atmospheres: results from a consistently coupled chemistry-radiation-hydrodynamics model},
  author={Drummond, Benjamin and H{\'e}brard, Eric and Mayne, Nathan J and Venot, Olivia and Ridgway, Robert J and Changeat, Quentin and Tsai, Shang-Min and Manners, James and Tremblin, Pascal and Abraham, Nathan Luke and others},
  journal={Astronomy \& Astrophysics},
  volume={636},
  pages={A68},
  year={2020},
  doi={https://doi.org/10.1051/0004-6361/201937153},
  publisher={EDP Sciences}
}

@article{mendoncca2018,
  title={Three-dimensional circulation driving chemical disequilibrium in WASP-43b},
  author={Mendon{\c{c}}a, Jo{\~a}o M and Tsai, Shang-min and Malik, Matej and Grimm, Simon L and Heng, Kevin},
  journal={The Astrophysical Journal},
  volume={869},
  number={2},
  pages={107},
  year={2018},
  DOI={10.3847/1538-4357/aaed23},
  publisher={The American Astronomical Society}
}

@article{drummond2016,
  title={The effects of consistent chemical kinetics calculations on the pressure-temperature profiles and emission spectra of hot Jupiters},
  author={Drummond, Benjamin and Tremblin, Pascal and Baraffe, Isabelle and Amundsen, David S and Mayne, Nathan J and Venot, Olivia and Goyal, Jayesh},
  journal={Astronomy \& Astrophysics},
  volume={594},
  pages={A69},
  year={2016},
  doi={https://doi.org/10.1051/0004-6361/201628799},
  publisher={EDP Sciences}
}

@article{mukherjee2023,
  title={PICASO 3.0: a one-dimensional climate model for giant planets and brown dwarfs},
  author={Mukherjee, Sagnick and Batalha, Natasha E and Fortney, Jonathan J and Marley, Mark S},
  journal={The Astrophysical Journal},
  volume={942},
  number={2},
  pages={71},
  year={2023},
  DOI={10.3847/1538-4357/ac9f48},
  publisher={The American Astronomical Society}
}

@article{mukherjee2024,
  title={The Sonora substellar atmosphere models. IV. Elf Owl: atmospheric mixing and chemical disequilibrium with varying metallicity and C/O ratios},
  author={Mukherjee, Sagnick and Fortney, Jonathan J and Morley, Caroline V and Batalha, Natasha E and Marley, Mark S and Karalidi, Theodora and Visscher, Channon and Lupu, Roxana and Freedman, Richard and Gharib-Nezhad, Ehsan},
  journal={The Astrophysical Journal},
  volume={963},
  number={1},
  pages={73},
  year={2024},
  DOI={10.3847/1538-4357/ad18c2},
  publisher={The American Astronomical Society}
}

@article{mukherjee2025,
  title={Effects of Planetary Parameters on Disequilibrium Chemistry in Irradiated Planetary Atmospheres: From Gas Giants to Sub-Neptunes},
  author={Mukherjee, Sagnick and Fortney, Jonathan J and Wogan, Nicholas F and Sing, David K and Ohno, Kazumasa},
  journal={The Astrophysical Journal},
  volume={985},
  number={2},
  pages={209},
  year={2025},
  DOI={10.3847/1538-4357/adc7b3},
  publisher={The American Astronomical Society}
}

@article{kirkpatrick2021a,
  title={The field substellar mass function based on the full-sky 20 pc census of 525 L, T, and Y dwarfs},
  author={Kirkpatrick, J Davy and Gelino, Christopher R and Faherty, Jacqueline K and Meisner, Aaron M and Caselden, Dan and Schneider, Adam C and Marocco, Federico and Cayago, Alfred J and Smart, RL and Eisenhardt, Peter R and others},
  journal={The Astrophysical Journal Supplement Series},
  volume={253},
  number={1},
  pages={7},
  year={2021},
  DOI={10.3847/1538-4365/abd107},
  publisher={The American Astronomical Society}
}

@article{lee2023dynamically,
  title={Dynamically coupled kinetic chemistry in brown dwarf atmospheres--I. Performing global scale kinetic modelling},
  author={Lee, Elspeth KH and Tan, Xianyu and Tsai, Shang-Min},
  journal={Monthly Notices of the Royal Astronomical Society},
  volume={523},
  number={3},
  pages={4477--4491},
  year={2023b},
  doi={https://doi.org/10.1093/mnras/stad1715},
  publisher={Oxford University Press}
}

@article{politano2004,
  title={The formation of cataclysmic variables with brown dwarf secondaries},
  author={Politano, Michael},
  journal={The Astrophysical Journal},
  volume={604},
  number={2},
  pages={817--826},
  DOI={10.1086/381958},
  year={2004}
}

@article{agundez2025,
  title={The mutual influence of disequilibrium composition and temperature in exoplanet atmospheres},
  author={Ag{\'u}ndez, Marcelino},
  journal={Astronomy \& Astrophysics},
  volume={699},
  pages={A306},
  year={2025},
  doi={https://doi.org/10.1051/0004-6361/202554732},
  publisher={EDP Sciences}
}

@article{biller2017,
  title={The time domain for brown dwarfs and directly imaged giant exoplanets: the power of variability monitoring},
  author={Biller, Beth},
  journal={Astronomical Review},
  volume={13},
  number={1},
  pages={1--27},
  year={2017},
  doi={https://doi.org/10.1080/21672857.2017.1303105},
  publisher={Taylor \& Francis}
}

@article{radigan2014,
  title={Strong brightness variations signal cloudy-to-clear transition of brown dwarfs},
  author={Radigan, Jacqueline and Lafreni{\`e}re, David and Jayawardhana, Ray and Artigau, Etienne},
  journal={The Astrophysical Journal},
  volume={793},
  number={2},
  pages={75},
  year={2014},
  DOI={10.1088/0004-637X/793/2/75},
  publisher={The American Astronomical Society}
}

@article{apai2013,
  title={HST spectral mapping of L/T transition brown dwarfs reveals cloud thickness variations},
  author={Apai, D{\'a}niel and Radigan, Jacqueline and Buenzli, Esther and Burrows, Adam and Reid, Iain Neill and Jayawardhana, Ray},
  journal={The Astrophysical Journal},
  volume={768},
  number={2},
  pages={121},
  year={2013},
  DOI={10.1088/0004-637X/768/2/121},
  publisher={The American Astronomical Society}
}

@article{buenzli2015,
  title={Cloud structure of the nearest brown dwarfs: Spectroscopic variability of Luhman 16AB from the Hubble Space Telescope},
  author={Buenzli, Esther and Saumon, Didier and Marley, Mark S and Apai, D{\'a}niel and Radigan, Jacqueline and Bedin, Luigi R and Reid, I Neill and Morley, Caroline V},
  journal={The Astrophysical Journal},
  volume={798},
  number={2},
  pages={127},
  year={2015},
  DOI={10.1088/0004-637X/798/2/127},
  publisher={The American Astronomical Society}
}

@article{lodders2002,
  title={Atmospheric chemistry in giant planets, brown dwarfs, and low-mass dwarf stars: I. Carbon, nitrogen, and oxygen},
  author={Lodders, Katharina and Fegley, Bruce},
  journal={Icarus},
  volume={155},
  number={2},
  pages={393--424},
  year={2002},
  doi={https://doi.org/10.1006/icar.2001.6740},
  publisher={Elsevier}
}

@incollection{lodders2006,
  title={Chemistry of low mass substellar objects},
  author={Lodders, Katharina and Fegley Jr, Bruce},
  booktitle={Astrophysics Update 2},
  pages={1--28},
  year={2006},
  DOI={https://doi.org/10.1007/3-540-30313-8_1},
  publisher={Springer}
}

@article{casewell2020,
  title={WD1032+ 011, an inflated brown dwarf in an old eclipsing binary with a white dwarf},
  author={Casewell, SL and Belardi, C and Parsons, SG and Littlefair, SP and Braker, IP and Hermes, JJ and Debes, John and Vanderbosch, Z and Burleigh, MR and G{\"a}nsicke, BT and others},
  journal={Monthly Notices of the Royal Astronomical Society},
  volume={497},
  number={3},
  pages={3571--3580},
  year={2020},
  doi={https://doi.org/10.1093/mnras/staa1608},
  publisher={Oxford University Press}
}

@article{lee2024,
  title={Dynamically coupled kinetic chemistry in brown dwarf atmospheres--II. Cloud and chemistry connections in directly imaged sub-Jupiter exoplanets},
  author={Lee, Elspeth KH and Tan, Xianyu and Tsai, Shang-Min},
  journal={Monthly Notices of the Royal Astronomical Society},
  volume={529},
  number={3},
  pages={2686--2701},
  year={2024},
  url={https://doi.org/10.1093/mnras/stae537},
  publisher={Oxford University Press}
}

@article{tsai2022,
  title={A mini-chemical scheme with net reactions for 3D general circulation models-I. Thermochemical kinetics},
  author={Tsai, Shang-Min and Lee, Elspeth KH and Pierrehumbert, Raymond},
  journal={Astronomy \& Astrophysics},
  volume={664},
  pages={A82},
  year={2022},
  url={https://doi.org/10.1051/0004-6361/202142816},
  publisher={EDP Sciences}
}

@article{lee2023,
  title={A mini-chemical scheme with net reactions for 3D general circulation models-II. 3D thermochemical modelling of WASP-39b and HD 189733b},
  author={Lee, Elspeth KH and Tsai, Shang-Min and Hammond, Mark and Tan, Xianyu},
  journal={Astronomy \& Astrophysics},
  volume={672},
  pages={A110},
  year={2023},
  url={https://doi.org/10.1051/0004-6361/202245473},
  publisher={EDP Sciences}
}

@article{tsai2017,
  title={VULCAN: an open-source, validated chemical kinetics Python code for exoplanetary atmospheres},
  author={Tsai, Shang-Min and Lyons, James R and Grosheintz, Luc and Rimmer, Paul B and Kitzmann, Daniel and Heng, Kevin},
  journal={The Astrophysical Journal Supplement Series},
  volume={228},
  number={2},
  pages={20},
  year={2017},
  DOI={10.3847/1538-4365/228/2/20},
  publisher={IOP Publishing}
}

@article{tsai2021,
  title={A comparative study of atmospheric chemistry with VULCAN},
  author={Tsai, Shang-Min and Malik, Matej and Kitzmann, Daniel and Lyons, James R and Fateev, Alexander and Lee, Elspeth and Heng, Kevin},
  journal={The Astrophysical Journal},
  volume={923},
  number={2},
  pages={264},
  year={2021},
  DOI={10.3847/1538-4357/ac29bc},
  publisher={IOP Publishing}
}

@article{Gierasch1968,
title = {A study of the thermal and dynamical structure of the martian lower atmosphere},
journal = {Planetary and Space Science},
volume = {16},
number = {5},
pages = {615-646},
year = {1968},
issn = {0032-0633},
doi = {https://doi.org/10.1016/0032-0633(68)90102-5},
url = {https://www.sciencedirect.com/science/article/pii/0032063368901025},
author = {Peter Gierasch and Richard Goody}
}

@article{Shapiro1970,
author = {Shapiro, Ralph},
title = {Smoothing, filtering, and boundary effects},
journal = {Reviews of Geophysics},
volume = {8},
number = {2},
pages = {359-387},
doi = {https://doi.org/10.1029/RG008i002p00359},
url = {https://agupubs.onlinelibrary.wiley.com/doi/abs/10.1029/RG008i002p00359},
eprint = {https://agupubs.onlinelibrary.wiley.com/doi/pdf/10.1029/RG008i002p00359},
year = {1970}
}

@ARTICLE{Teinturier2026,
       author = {{Teinturier}, Lucas and {Charnay}, Benjamin and {Spiga}, Aymeric and {B{\'e}zard}, Bruno},
        title = "{Clouds as the driver of variability and colour changes in brown dwarf atmospheres}",
      journal = {Nature Astronomy},
         year = 2026,
        month = feb,
       volume = {10},
        pages = {224-233},
          doi = {10.1038/s41550-025-02709-1},
archivePrefix = {arXiv},
       eprint = {2510.18090},
 primaryClass = {astro-ph.EP},
       adsurl = {https://ui.adsabs.harvard.edu/abs/2026NatAs..10..224T}
}

@ARTICLE{tan2025,
       author = {{Tan}, Xianyu and {Zhang}, Xi and {Marley}, Mark S. and {Zhou}, Yifan and {Lew}, Ben W.~P. and {Miles}, Brittany E. and {Batalha}, Natasha E. and {Biller}, Beth A. and {Chauvin}, Ga{\"e}l and {Hinkley}, Sasha and {Hoch}, Kielan K.~W. and {Manjavacas}, Elena and {Metchev}, Stanimir and {Petrus}, Simon and {Rickman}, Emily and {Skemer}, Andrew and {Su{\'a}rez}, Genaro and {Sutlieff}, Ben J. and {Vos}, Johanna M. and {Whiteford}, Niall},
        title = "{Large-amplitude variability driven by giant dust storms on a planetary-mass companion}",
      journal = {Science Advances},
         year = 2025,
        month = nov,
       volume = {11},
        pages = {22.3324},
          doi = {10.1126/sciadv.adv3324},
archivePrefix = {arXiv},
       eprint = {2511.23163},
 primaryClass = {astro-ph.EP},
       adsurl = {https://ui.adsabs.harvard.edu/abs/2025SciA...11v3324T}
}

@ARTICLE{lai2026,
       author = {{Lai}, Yanhong and {Tan}, Xianyu and {Su}, Yubo},
        title = "{Atmospheric Circulation of High-obliquity Mini-Neptunes}",
      journal = {\apj},
         year = 2026,
        month = mar,
       volume = {999},
       number = {1},
          eid = {36},
        pages = {36},
          doi = {10.3847/1538-4357/ae3d35},
archivePrefix = {arXiv},
       eprint = {2601.18606},
 primaryClass = {astro-ph.EP},
       adsurl = {https://ui.adsabs.harvard.edu/abs/2026ApJ...999...36L}
}

@ARTICLE{tan2022bd,
       author = {{Tan}, Xianyu},
        title = "{Jet streams and tracer mixing in the atmospheres of brown dwarfs and isolated young giant planets}",
      journal = {\mnras},
         year = 2022,
        month = apr,
       volume = {511},
       number = {4},
        pages = {4861-4881},
          doi = {10.1093/mnras/stac344},
archivePrefix = {arXiv},
       eprint = {2203.10523},
 primaryClass = {astro-ph.EP},
       adsurl = {https://ui.adsabs.harvard.edu/abs/2022MNRAS.511.4861T}
}

@book{showman2010,
  author    = {Showman, Adam P. and Cho, James Y-K. and Menou, Kristian},
  title     = {Atmospheric Circulation of Exoplanets},
  year      = {2010},
  publisher = {University of Arizona Press},
  address   = {Tucson, AZ},
  series    = {Exoplanets Series},
  isbn      = {978-0-8165-3059-5}
}

@article{knutson2014,
  title={A featureless transmission spectrum for the Neptune-mass exoplanet GJ 436b},
  author={Knutson, Heather A and Benneke, Bj{\"o}rn and Deming, Drake and Homeier, Derek},
  journal={Nature},
  volume={505},
  number={7481},
  pages={66--68},
  year={2014},
  url={https://doi.org/10.1038/nature12887},
  publisher={Nature Publishing Group UK London}
}

@article{demory2013,
  title={Inference of inhomogeneous clouds in an exoplanet atmosphere},
  author={Demory, Brice-Olivier and De Wit, Julien and Lewis, Nikole and Fortney, Jonathan and Zsom, Andras and Seager, Sara and Knutson, Heather and Heng, Kevin and Madhusudhan, Nikku and Gillon, Michael and others},
  journal={The Astrophysical Journal Letters},
  volume={776},
  number={2},
  pages={L25},
  year={2013},
  DOI={10.1088/2041-8205/776/2/L25},
  publisher={IOP Publishing}
}

@article{sing2016,
  title={A continuum from clear to cloudy hot-Jupiter exoplanets without primordial water depletion},
  author={Sing, David K and Fortney, Jonathan J and Nikolov, Nikolay and Wakeford, Hannah R and Kataria, Tiffany and Evans, Thomas M and Aigrain, Suzanne and Ballester, Gilda E and Burrows, Adam S and Deming, Drake and others},
  journal={Nature},
  volume={529},
  number={7584},
  pages={59--62},
  year={2016},
  url={https://doi.org/10.1038/nature16068},
  publisher={Nature Publishing Group UK London}
}

@article{feinstein2023,
  title={Early Release Science of the exoplanet WASP-39b with JWST NIRISS},
  author={Feinstein, Adina D and Radica, Michael and Welbanks, Luis and Murray, Catriona Anne and Ohno, Kazumasa and Coulombe, Louis-Philippe and Espinoza, N{\'e}stor and Bean, Jacob L and Teske, Johanna K and Benneke, Bj{\"o}rn and others},
  journal={Nature},
  volume={614},
  number={7949},
  pages={670--675},
  year={2023},
  url={https://doi.org/10.1038/s41586-022-05674-1},
  publisher={Nature Publishing Group UK London}
}

@article{showman2016,
  title={Illuminating brown dwarfs},
  author={Showman, Adam P},
  journal={Nature},
  volume={533},
  number={7603},
  pages={330--331},
  year={2016},
  DOI={https://doi.org/10.1038/533330a},
  publisher={Nature Publishing Group UK London}
}

@article{parsons2025,
  title={Two almost planetary mass survivors of common envelope evolution},
  author={Parsons, Steven G and Brown, Alex J and Casewell, SL and Littlefair, Stuart P and van Roestel, Jan and Rebassa-Mansergas, Alberto and Murillo-Ojeda, Raquel and Hollands, MA and Zorotovic, M{\'o}nica and Castro Segura, Noel and others},
  journal={Monthly Notices of the Royal Astronomical Society},
  volume={537},
  number={2},
  pages={2112--2126},
  year={2025},
  url={https://doi.org/10.1093/mnras/staf166},
  publisher={Oxford University Press}
}

@article{dewit2025,
  title={An Ultra-hot Low-mass Brown Dwarf Eclipsing the Puffy White Dwarf J0404+ 1112},
  author={de Wit, J and Alonso, R and Rackham, BV and Lederer, SM and Burdanov, AY and Barkaoui, K and Albornoz, M and Cavalier, T and Gillon, M and Triaud, AHMJ and others},
  journal={Research Notes of the AAS},
  volume={9},
  number={10},
  pages={270},
  year={2025},
  DOI={10.3847/2515-5172/ae1228},
  publisher={The American Astronomical Society}
}

@ARTICLE{tan2020,
       author = {{Tan}, Xianyu and {Showman}, Adam P.},
        title = "{Atmospheric Circulation of Tidally Locked Gas Giants with Increasing Rotation and Implications for White Dwarf-Brown Dwarf Systems}",
      journal = {\apj},
         year = 2020,
        month = oct,
       volume = {902},
       number = {1},
          eid = {27},
        pages = {27},
          doi = {10.3847/1538-4357/abb3d4},
archivePrefix = {arXiv},
       eprint = {2001.06269},
 primaryClass = {astro-ph.EP},
       adsurl = {https://ui.adsabs.harvard.edu/abs/2020ApJ...902...27T}
}

@article{showman2013,
  title={Atmospheric circulation of terrestrial exoplanets},
  author={Showman, Adam P and Wordsworth, Robin D and Merlis, Timothy M and Kaspi, Yohai},
  journal={Comparative climatology of terrestrial planets},
  volume={1},
  pages={277--326},
  year={2013},
  doi = {10.2458/azu_uapress_9780816530595-ch012},
  adsurl = {https://ui.adsabs.harvard.edu/abs/2013cctp.book..277S},
  publisher={University of Arizona Press, Tuscon, AZ}
}

@ARTICLE{tan2024,
       author = {{Tan}, Xianyu and {Komacek}, Thaddeus D. and {Batalha}, Natasha E. and {Deming}, Drake and {Lupu}, Roxana and {Parmentier}, Vivien and {Pierrehumbert}, Raymond T.},
        title = "{Modelling the day-night temperature variations of ultra-hot Jupiters: confronting non-grey general circulation models and observations}",
      journal = {\mnras},
         year = 2024,
        month = feb,
       volume = {528},
       number = {1},
        pages = {1016-1036},
          doi = {10.1093/mnras/stae050},
archivePrefix = {arXiv},
       eprint = {2401.03859},
 primaryClass = {astro-ph.EP},
       adsurl = {https://ui.adsabs.harvard.edu/abs/2024MNRAS.528.1016T}
}

@article{blumenthal2018,
  title={A comparison of simulated JWST observations derived from equilibrium and non-equilibrium chemistry models of giant exoplanets},
  author={Blumenthal, Sarah D and Mandell, Avi M and H{\'e}brard, Eric and Batalha, Natasha E and Cubillos, Patricio E and Rugheimer, Sarah and Wakeford, Hannah R},
  journal={The Astrophysical Journal},
  volume={853},
  number={2},
  pages={138},
  year={2018},
  DOI={10.3847/1538-4357/aa9e51},
  publisher={The American Astronomical Society}
}

@article{tsai2023,
  title={Photochemically produced SO2 in the atmosphere of WASP-39b},
  author={Tsai, Shang-Min and Lee, Elspeth KH and Powell, Diana and Gao, Peter and Zhang, Xi and Moses, Julianne and H{\'e}brard, Eric and Venot, Olivia and Parmentier, Vivien and Jordan, Sean and others},
  journal={Nature},
  volume={617},
  number={7961},
  pages={483--487},
  year={2023},
  url={https://doi.org/10.1038/s41586-023-05902-2},
  publisher={Nature Publishing Group UK London}
}

@article{roman2021,
  title={Clouds in three-dimensional models of hot Jupiters over a wide range of temperatures. I. Thermal structures and broadband phase-curve predictions},
  author={Roman, Michael T and Kempton, Eliza M-R and Rauscher, Emily and Harada, Caleb K and Bean, Jacob L and Stevenson, Kevin B},
  journal={The Astrophysical Journal},
  volume={908},
  number={1},
  pages={101},
  year={2021},
  url={https://doi.org/10.3847/1538-4357/abd549},
  publisher={The American Astronomical Society}
}

@article{roman2019,
  title={Modeled temperature-dependent clouds with radiative feedback in hot Jupiter atmospheres},
  author={Roman, Michael and Rauscher, Emily},
  journal={The Astrophysical Journal},
  volume={872},
  number={1},
  pages={1},
  year={2019},
  url={https://doi.org/10.3847/1538-4357/aafdb5},
  publisher={The American Astronomical Society}
}

@article{beatty2019,
  title={Spitzer phase curves of KELT-1b and the signatures of nightside clouds in thermal phase observations},
  author={Beatty, Thomas G and Marley, Mark S and Gaudi, B Scott and Col{\'o}n, Knicole D and Fortney, Jonathan J and Showman, Adam P},
  journal={The Astronomical Journal},
  volume={158},
  number={4},
  pages={166},
  year={2019},
  DOI={10.3847/1538-3881/ab33fc},
  publisher={The American Astronomical Society}
}

@article{sing2024,
  title={A warm Neptune’s methane reveals core mass and vigorous atmospheric mixing},
  author={Sing, David K and Rustamkulov, Zafar and Thorngren, Daniel P and Barstow, Joanna K and Tremblin, Pascal and Alves de Oliveira, Catarina and Beck, Tracy L and Birkmann, Stephan M and Challener, Ryan C and Crouzet, Nicolas and others},
  journal={Nature},
  volume={630},
  number={8018},
  pages={831--835},
  year={2024},
  url={https://doi.org/10.1038/s41586-024-07395-z},
  publisher={Nature Publishing Group UK London}
}

@ARTICLE{bell2024,
       author = {{Bell}, Taylor J. and {Crouzet}, Nicolas and {Cubillos}, Patricio E. and {Kreidberg}, Laura and {Piette}, Anjali A.~A. and {Roman}, Michael T. and {Barstow}, Joanna K. and {Blecic}, Jasmina and {Carone}, Ludmila and {Coulombe}, Louis-Philippe and {Ducrot}, Elsa and {Hammond}, Mark and {Mendon{\c{c}}a}, Jo{\~a}o M. and {Moses}, Julianne I. and {Parmentier}, Vivien and {Stevenson}, Kevin B. and {Teinturier}, Lucas and {Zhang}, Michael and {Batalha}, Natalie M. and {Bean}, Jacob L. and {Benneke}, Bj{\"o}rn and {Charnay}, Benjamin and {Chubb}, Katy L. and {Demory}, Brice-Olivier and {Gao}, Peter and {Lee}, Elspeth K.~H. and {L{\'o}pez-Morales}, Mercedes and {Morello}, Giuseppe and {Rauscher}, Emily and {Sing}, David K. and {Tan}, Xianyu and {Venot}, Olivia and {Wakeford}, Hannah R. and {Aggarwal}, Keshav and {Ahrer}, Eva-Maria and {Alam}, Munazza K. and {Baeyens}, Robin and {Barrado}, David and {Caceres}, Claudio and {Carter}, Aarynn L. and {Casewell}, Sarah L. and {Challener}, Ryan C. and {Crossfield}, Ian J.~M. and {Decin}, Leen and {D{\'e}sert}, Jean-Michel and {Dobbs-Dixon}, Ian and {Dyrek}, Achr{\`e}ne and {Espinoza}, N{\'e}stor and {Feinstein}, Adina D. and {Gibson}, Neale P. and {Harrington}, Joseph and {Helling}, Christiane and {Hu}, Renyu and {Iro}, Nicolas and {Kempton}, Eliza M. -R. and {Kendrew}, Sarah and {Komacek}, Thaddeus D. and {Krick}, Jessica and {Lagage}, Pierre-Olivier and {Leconte}, J{\'e}r{\'e}my and {Lendl}, Monika and {Lewis}, Neil T. and {Lothringer}, Joshua D. and {Malsky}, Isaac and {Mancini}, Luigi and {Mansfield}, Megan and {Mayne}, Nathan J. and {Evans-Soma}, Thomas M. and {Molaverdikhani}, Karan and {Nikolov}, Nikolay K. and {Nixon}, Matthew C. and {Palle}, Enric and {Petit dit de la Roche}, Dominique J.~M. and {Piaulet}, Caroline and {Powell}, Diana and {Rackham}, Benjamin V. and {Schneider}, Aaron D. and {Steinrueck}, Maria E. and {Taylor}, Jake and {Welbanks}, Luis and {Yurchenko}, Sergei N. and {Zhang}, Xi and {Zieba}, Sebastian},
        title = "{Nightside clouds and disequilibrium chemistry on the hot Jupiter WASP-43b}",
      journal = {Nature Astronomy},
         year = 2024,
        month = jul,
       volume = {8},
        pages = {879-898},
          doi = {10.1038/s41550-024-02230-x},
archivePrefix = {arXiv},
       eprint = {2401.13027},
 primaryClass = {astro-ph.EP},
       adsurl = {https://ui.adsabs.harvard.edu/abs/2024NatAs...8..879B}
}

@article{lodders2003,
  title={Solar system abundances and condensation temperatures of the elements},
  author={Lodders, Katharina},
  journal={The Astrophysical Journal},
  volume={591},
  number={2},
  pages={1220},
  year={2003},
  DOI={10.1086/375492},
  publisher={IOP Publishing}
}

@ARTICLE{komacek2019mixing,
       author = {{Komacek}, Thaddeus D. and {Showman}, Adam P. and {Parmentier}, Vivien},
        title = "{Vertical Tracer Mixing in Hot Jupiter Atmospheres}",
      journal = {\apj},
         year = 2019,
        month = aug,
       volume = {881},
       number = {2},
          eid = {152},
        pages = {152},
          doi = {10.3847/1538-4357/ab338b},
archivePrefix = {arXiv},
       eprint = {1904.09676},
 primaryClass = {astro-ph.EP},
       adsurl = {https://ui.adsabs.harvard.edu/abs/2019ApJ...881..152K}
}

@ARTICLE{komacek2022cloud,
       author = {{Komacek}, Thaddeus D. and {Tan}, Xianyu and {Gao}, Peter and {Lee}, Elspeth K.~H.},
        title = "{Patchy Nightside Clouds on Ultra-hot Jupiters: General Circulation Model Simulations with Radiatively Active Cloud Tracers}",
      journal = {\apj},
         year = 2022,
        month = jul,
       volume = {934},
       number = {1},
          eid = {79},
        pages = {79},
          doi = {10.3847/1538-4357/ac7723},
archivePrefix = {arXiv},
       eprint = {2205.07834},
 primaryClass = {astro-ph.EP},
       adsurl = {https://ui.adsabs.harvard.edu/abs/2022ApJ...934...79K}
}

@ARTICLE{tan2019uhj,
       author = {{Tan}, Xianyu and {Komacek}, Thaddeus D.},
        title = "{The Atmospheric Circulation of Ultra-hot Jupiters}",
      journal = {\apj},
         year = 2019,
        month = nov,
       volume = {886},
       number = {1},
          eid = {26},
        pages = {26},
          doi = {10.3847/1538-4357/ab4a76},
archivePrefix = {arXiv},
       eprint = {1910.01622},
 primaryClass = {astro-ph.EP},
       adsurl = {https://ui.adsabs.harvard.edu/abs/2019ApJ...886...26T}
}

@ARTICLE{komacek2017,
       author = {{Komacek}, Thaddeus D. and {Showman}, Adam P. and {Tan}, Xianyu},
        title = "{Atmospheric Circulation of Hot Jupiters: Dayside-Nightside Temperature Differences. II. Comparison with Observations}",
      journal = {\apj},
         year = 2017,
        month = feb,
       volume = {835},
       number = {2},
          eid = {198},
        pages = {198},
          doi = {10.3847/1538-4357/835/2/198},
archivePrefix = {arXiv},
       eprint = {1610.03893},
 primaryClass = {astro-ph.EP},
       adsurl = {https://ui.adsabs.harvard.edu/abs/2017ApJ...835..198K}
}

@ARTICLE{mehta2025,
       author = {{Mehta}, Nishil and {Parmentier}, Vivien and {Tan}, Xianyu and {Lee}, Elspeth K.~H. and {Guillot}, Tristan and {Wiser}, Lindsey S. and {Bell}, Taylor J. and {Schlawin}, Everett and {Arnold}, Kenneth and {Mukherjee}, Sagnick and {Greene}, Thomas P. and {Beatty}, Thomas G. and {Welbanks}, Luis and {Line}, Michael R. and {Murphy}, Matthew M. and {Fortney}, Jonathan J. and {Ohno}, Kazumasa},
        title = "{How clear are the skies of WASP-80b?: 3D Cloud feedback on the atmosphere and spectra of the warm Jupiter}",
      journal = {arXiv e-prints},
         year = 2025,
        month = sep,
          eid = {arXiv:2509.23406},
        pages = {arXiv:2509.23406},
          doi = {10.48550/arXiv.2509.23406},
archivePrefix = {arXiv},
       eprint = {2509.23406},
 primaryClass = {astro-ph.EP},
       adsurl = {https://ui.adsabs.harvard.edu/abs/2025arXiv250923406M}
}

@article{steinrueck2023,
  title={Photochemical hazes dramatically alter temperature structure and atmospheric circulation in 3D simulations of hot Jupiters},
  author={Steinrueck, Maria E and Koskinen, Tommi and Lavvas, Panayotis and Parmentier, Vivien and Zieba, Sebastian and Tan, Xianyu and Zhang, Xi and Kreidberg, Laura},
  journal={The Astrophysical Journal},
  volume={951},
  number={2},
  pages={117},
  year={2023},
  DOI={10.3847/1538-4357/acd4bb},
  publisher={IOP Publishing}
}

@article{adcroft2004,
  title={Implementation of an atmosphere--ocean general circulation model on the expanded spherical cube},
  author={Adcroft, Alistair and Campin, Jean-Michel and Hill, Chris and Marshall, John},
  journal={Monthly Weather Review},
  volume={132},
  number={12},
  pages={2845--2863},
  DOI={https://doi.org/10.1175/MWR2823.1},
  year={2004}
}

@ARTICLE{showman2020review,
       author = {{Showman}, Adam P. and {Tan}, Xianyu and {Parmentier}, Vivien},
        title = "{Atmospheric Dynamics of Hot Giant Planets and Brown Dwarfs}",
      journal = {\ssr},
         year = 2020,
        month = dec,
       volume = {216},
       number = {8},
          eid = {139},
        pages = {139},
          doi = {10.1007/s11214-020-00758-8},
archivePrefix = {arXiv},
       eprint = {2007.15363},
 primaryClass = {astro-ph.EP},
       adsurl = {https://ui.adsabs.harvard.edu/abs/2020SSRv..216..139S}
}

@article{gao2020,
  title={Aerosol composition of hot giant exoplanets dominated by silicates and hydrocarbon hazes},
  author={Gao, Peter and Thorngren, Daniel P and Lee, Elspeth KH and Fortney, Jonathan J and Morley, Caroline V and Wakeford, Hannah R and Powell, Diana K and Stevenson, Kevin B and Zhang, Xi},
  journal={Nature Astronomy},
  volume={4},
  number={10},
  pages={951--956},
  year={2020},
  url={https://doi.org/10.1038/s41550-020-1114-3},
  publisher={Nature Publishing Group UK London}
}

@article{showman2009,
  title={Atmospheric circulation of hot Jupiters: Coupled radiative-dynamical general circulation model simulations of HD 189733b and HD 209458b},
  author={Showman, Adam P and Fortney, Jonathan J and Lian, Yuan and Marley, Mark S and Freedman, Richard S and Knutson, Heather A and Charbonneau, David},
  journal={The Astrophysical Journal},
  volume={699},
  number={1},
  pages={564},
  year={2009},
  DOI={10.1088/0004-637X/699/1/564},
  publisher={IOP Publishing}
}

@article{marley1999,
  title={Reflected spectra and albedos of extrasolar giant planets. I. Clear and cloudy atmospheres},
  author={Marley, Mark S and Gelino, Christopher and Stephens, Denise and Lunine, Jonathan I and Freedman, Richard},
  journal={The Astrophysical Journal},
  volume={513},
  number={2},
  pages={879},
  year={1999},
  DOI={10.1086/306881},
  publisher={IOP Publishing}
}

@article{parmentier2013,
  title={3d mixing in hot jupiters atmospheres-i. application to the day/night cold trap in hd 209458b},
  author={Parmentier, Vivien and Showman, Adam P and Lian, Yuan},
  journal={Astronomy \& Astrophysics},
  volume={558},
  pages={A91},
  year={2013},
  url={https://doi.org/10.1051/0004-6361/201321132},
  publisher={EDP Sciences}
}

@article{visscher2010,
  title={Atmospheric chemistry in giant planets, brown dwarfs, and low-mass dwarf stars. III. Iron, magnesium, and silicon},
  author={Visscher, Channon and Lodders, Katharina and Fegley, Bruce},
  journal={The Astrophysical Journal},
  volume={716},
  number={2},
  pages={1060},
  year={2010},
  DOI={10.1088/0004-637X/716/2/1060},
  publisher={IOP Publishing}
}

@article{batalha2019,
  title={Exoplanet reflected-light spectroscopy with PICASO},
  author={Batalha, Natasha E and Marley, Mark S and Lewis, Nikole K and Fortney, Jonathan J},
  journal={The Astrophysical Journal},
  volume={878},
  number={1},
  pages={70},
  year={2019},
  DOI={10.3847/1538-4357/ab1b51},
  publisher={IOP Publishing}
}

@article{showman2011,
  title={Equatorial superrotation on tidally locked exoplanets},
  author={Showman, Adam P and Polvani, Lorenzo M},
  journal={The Astrophysical Journal},
  volume={738},
  number={1},
  pages={71},
  year={2011},
  doi={10.1088/0004-637X/738/1/71},
  publisher={IOP Publishing}
}

@article{pierrehumbert2019,
  title={Atmospheric circulation of tide-locked exoplanets},
  author={Pierrehumbert, Raymond T and Hammond, Mark},
  journal={Annual Review of Fluid Mechanics},
  volume={51},
  number={1},
  pages={275--303},
  year={2019},
  url={https://doi.org/10.1146/annurev- fluid- 010518- 040516},
  publisher={Annual Reviews}
}

@article{kataria2013,
  title={Three-dimensional atmospheric circulation of hot Jupiters on highly eccentric orbits},
  author={Kataria, Tiffany and Showman, Adam P and Lewis, Nikole K and Fortney, Jonathan J and Marley, Mark S and Freedman, Richard S},
  journal={The Astrophysical Journal},
  volume={767},
  number={1},
  pages={76},
  year={2013},
  doi={10.1088/0004-637X/767/1/76},
  publisher={IOP Publishing}
}

@ARTICLE{Amaro2025,
       author = {{Amaro}, Rachael C. and {Apai}, D{\'a}niel and {Zhou}, Yifan and {Lothringer}, Joshua D. and {Casewell}, Sarah L. and {Tan}, Xianyu and {Lew}, Ben W.~P. and {Barman}, Travis and {Marley}, Mark S. and {Mayorga}, L.~C. and {Parmentier}, Vivien},
        title = "{Phase-resolved Hubble Space Telescope WFC3 Spectroscopy of the Weakly Irradiated Brown Dwarf GD 1400 and Energy Redistribution─Irradiation Trends in Six White Dwarf─Brown Dwarf Binaries}",
      journal = {\apj},
         year = 2025,
        month = feb,
       volume = {979},
       number = {2},
          eid = {231},
        pages = {231},
          doi = {10.3847/1538-4357/ada295},
archivePrefix = {arXiv},
       eprint = {2501.05609},
 primaryClass = {astro-ph.SR},
       adsurl = {https://ui.adsabs.harvard.edu/abs/2025ApJ...979..231A}
}

@ARTICLE{French2024,
       author = {{French}, Jenni R. and {Casewell}, Sarah L. and {Amaro}, Rachael C. and {Lothringer}, Joshua D. and {Mayorga}, L.~C. and {Littlefair}, Stuart P. and {Lew}, Ben W.~P. and {Zhou}, Yifan and {Apai}, Daniel and {Marley}, Mark S. and {Parmentier}, Vivien and {Tan}, Xianyu},
        title = "{The only inflated brown dwarf in an eclipsing white dwarf-brown dwarf binary: WD1032+011B}",
      journal = {\mnras},
         year = 2024,
        month = nov,
       volume = {534},
       number = {3},
        pages = {2244-2262},
          doi = {10.1093/mnras/stae2121},
archivePrefix = {arXiv},
       eprint = {2409.06874},
 primaryClass = {astro-ph.SR},
       adsurl = {https://ui.adsabs.harvard.edu/abs/2024MNRAS.534.2244F}
}

@ARTICLE{Lothringer2024,
       author = {{Lothringer}, Joshua D. and {Zhou}, Yifan and {Apai}, D{\'a}niel and {Tan}, Xianyu and {Parmentier}, Vivien and {Casewell}, Sarah L.},
        title = "{Atmospheric Retrievals of the Phase-resolved Spectra of Irradiated Brown Dwarfs WD-0137B and EPIC-2122B}",
      journal = {\apj},
         year = 2024,
        month = jun,
       volume = {968},
       number = {2},
          eid = {126},
        pages = {126},
          doi = {10.3847/1538-4357/ad43da},
archivePrefix = {arXiv},
       eprint = {2404.16813},
 primaryClass = {astro-ph.SR},
       adsurl = {https://ui.adsabs.harvard.edu/abs/2024ApJ...968..126L}
}

@ARTICLE{Amaro2024,
       author = {{Amaro}, Rachael C. and {Apai}, D{\'a}niel and {Lew}, Ben W.~P. and {Zhou}, Yifan and {Lothringer}, Joshua D. and {Casewell}, Sarah L. and {Tan}, Xianyu and {Barman}, Travis and {Marley}, Mark S. and {Mayorga}, L.~C. and {Parmentier}, Vivien},
        title = "{Time-resolved Hubble Space Telescope Wide Field Camera 3 Spectrophotometry Reveals Inefficient Day-to-night Heat Redistribution in the Highly Irradiated Brown Dwarf SDSS 1557B}",
      journal = {\apj},
         year = 2024,
        month = may,
       volume = {966},
       number = {1},
          eid = {4},
        pages = {4},
          doi = {10.3847/1538-4357/ad354c},
archivePrefix = {arXiv},
       eprint = {2404.08087},
 primaryClass = {astro-ph.EP},
       adsurl = {https://ui.adsabs.harvard.edu/abs/2024ApJ...966....4A}
}

@ARTICLE{Amaro2023,
       author = {{Amaro}, Rachael C. and {Apai}, D{\'a}niel and {Zhou}, Yifan and {Lew}, Ben W.~P. and {Casewell}, Sarah L. and {Mayorga}, L.~C. and {Marley}, Mark S. and {Tan}, Xianyu and {Lothringer}, Joshua D. and {Parmentier}, Vivien and {Barman}, Travis},
        title = "{Hotter than Expected: Hubble Space Telescope (HST)/WFC3 Phase-resolved Spectroscopy of a Rare Irradiated Brown Dwarf with Strong Internal Heat Flux}",
      journal = {\apj},
         year = 2023,
        month = may,
       volume = {948},
       number = {2},
          eid = {129},
        pages = {129},
          doi = {10.3847/1538-4357/acbfb3},
archivePrefix = {arXiv},
       eprint = {2303.07420},
 primaryClass = {astro-ph.EP},
       adsurl = {https://ui.adsabs.harvard.edu/abs/2023ApJ...948..129A}
}

@ARTICLE{Zhou2022,
       author = {{Zhou}, Yifan and {Apai}, D{\'a}niel and {Tan}, Xianyu and {Lothringer}, Joshua D. and {Lew}, Ben W.~P. and {Casewell}, Sarah L. and {Parmentier}, Vivien and {Marley}, Mark S. and {Xu}, Siyi and {Mayorga}, L.~C.},
        title = "{HST/WFC3 Complete Phase-resolved Spectroscopy of White-dwarf-brown-dwarf Binaries WD 0137 and EPIC 2122}",
      journal = {\aj},
         year = 2022,
        month = jan,
       volume = {163},
       number = {1},
          eid = {17},
        pages = {17},
          doi = {10.3847/1538-3881/ac3095},
archivePrefix = {arXiv},
       eprint = {2110.10162},
 primaryClass = {astro-ph.SR},
       adsurl = {https://ui.adsabs.harvard.edu/abs/2022AJ....163...17Z}
}

@ARTICLE{grant2023,
       author = {{Grant}, David and {Lewis}, Nikole K. and {Wakeford}, Hannah R. and {Batalha}, Natasha E. and {Glidden}, Ana and {Goyal}, Jayesh and {Mullens}, Elijah and {MacDonald}, Ryan J. and {May}, Erin M. and {Seager}, Sara and {Stevenson}, Kevin B. and {Valenti}, Jeff A. and {Visscher}, Channon and {Alderson}, Lili and {Allen}, Natalie H. and {Ca{\~n}as}, Caleb I. and {Col{\'o}n}, Knicole and {Clampin}, Mark and {Espinoza}, N{\'e}stor and {Gressier}, Am{\'e}lie and {Huang}, Jingcheng and {Lin}, Zifan and {Long}, Douglas and {Louie}, Dana R. and {Pe{\~n}a-Guerrero}, Maria and {Ranjan}, Sukrit and {Sotzen}, Kristin S. and {Valentine}, Daniel and {Anderson}, Jay and {Balmer}, William O. and {Bellini}, Andrea and {Hoch}, Kielan K.~W. and {Kammerer}, Jens and {Libralato}, Mattia and {Mountain}, C. Matt and {Perrin}, Marshall D. and {Pueyo}, Laurent and {Rickman}, Emily and {Rebollido}, Isabel and {Sohn}, Sangmo Tony and {van der Marel}, Roeland P. and {Watkins}, Laura L.},
        title = "{JWST-TST DREAMS: Quartz Clouds in the Atmosphere of WASP-17b}",
      journal = {\apjl},
         year = 2023,
        month = oct,
       volume = {956},
       number = {2},
          eid = {L32},
        pages = {L32},
          doi = {10.3847/2041-8213/acfc3b10.3847/2041-8213/acfdab},
archivePrefix = {arXiv},
       eprint = {2310.08637},
 primaryClass = {astro-ph.EP},
       adsurl = {https://ui.adsabs.harvard.edu/abs/2023ApJ...956L..32G}
}

@ARTICLE{cushing2006,
       author = {{Cushing}, Michael C. and {Roellig}, Thomas L. and {Marley}, Mark S. and {Saumon}, D. and {Leggett}, S.~K. and {Kirkpatrick}, J. Davy and {Wilson}, John C. and {Sloan}, G.~C. and {Mainzer}, Amy K. and {Van Cleve}, Jeff E. and {Houck}, James R.},
        title = "{A Spitzer Infrared Spectrograph Spectral Sequence of M, L, and T Dwarfs}",
      journal = {\apj},
         year = 2006,
        month = sep,
       volume = {648},
       number = {1},
        pages = {614-628},
          doi = {10.1086/505637},
archivePrefix = {arXiv},
       eprint = {astro-ph/0605639},
 primaryClass = {astro-ph},
       adsurl = {https://ui.adsabs.harvard.edu/abs/2006ApJ...648..614C}
}

@ARTICLE{suarez2023,
       author = {{Su{\'a}rez}, Genaro and {Metchev}, Stanimir},
        title = "{Ultracool dwarfs observed with the Spitzer Infrared Spectrograph - III. Dust grains in young L dwarf atmospheres are heavier}",
      journal = {\mnras},
         year = 2023,
        month = aug,
       volume = {523},
       number = {3},
        pages = {4739-4747},
          doi = {10.1093/mnras/stad1711},
archivePrefix = {arXiv},
       eprint = {2306.01119},
 primaryClass = {astro-ph.SR},
       adsurl = {https://ui.adsabs.harvard.edu/abs/2023MNRAS.523.4739S}
}

@ARTICLE{fortney2020,
       author = {{Fortney}, Jonathan J. and {Visscher}, Channon and {Marley}, Mark S. and {Hood}, Callie E. and {Line}, Michael R. and {Thorngren}, Daniel P. and {Freedman}, Richard S. and {Lupu}, Roxana},
        title = "{Beyond Equilibrium Temperature: How the Atmosphere/Interior Connection Affects the Onset of Methane, Ammonia, and Clouds in Warm Transiting Giant Planets}",
      journal = {\aj},
         year = 2020,
        month = dec,
       volume = {160},
       number = {6},
          eid = {288},
        pages = {288},
          doi = {10.3847/1538-3881/abc5bd},
archivePrefix = {arXiv},
       eprint = {2010.00146},
 primaryClass = {astro-ph.EP},
       adsurl = {https://ui.adsabs.harvard.edu/abs/2020AJ....160..288F}
}

@MISC{zhou2024jwst,
       author = {{Zhou}, Yifan and {Amaro}, Rachael C. and {Apai}, Daniel and {Casewell}, Sarah and {French}, Jenni and {Lew}, Ben Wei Peng and {Lothringer}, Joshua D. and {Marley}, Mark S. and {Mayorga}, Laura C. and {Parmentier}, Vivien and {Tan}, Xianyu and {Xu}, Siyi},
        title = "{Dwarfs of Fire and Ice: Mapping the Irradiated Atmospheres of White Dwarf-Brown Dwarf Binaries with NIRSpec PRISM Phase Curves}",
 howpublished = {JWST Proposal. Cycle 3, ID. \#4967},
         year = 2024,
        month = feb,
        pages = {4967},
       adsurl = {https://ui.adsabs.harvard.edu/abs/2024jwst.prop.4967Z}
}

@article{marley2015,
  title={On the cool side: modeling the atmospheres of brown dwarfs and giant planets},
  author={Marley, Mark S and Robinson, Tyler D},
  journal={Annual Review of Astronomy and Astrophysics},
  volume={53},
  number={1},
  pages={279--323},
  year={2015},
  url={https://doi.org/10.1146/annurev-astro-082214-122522},
  publisher={Annual Reviews}
}

@ARTICLE{Lee2020wdbd,
       author = {{Lee}, Elspeth K.~H. and {Casewell}, Sarah L. and {Chubb}, Katy L. and {Hammond}, Mark and {Tan}, Xianyu and {Tsai}, Shang-Min and {Pierrehumbert}, Raymond T.},
        title = "{Simplified 3D GCM modelling of the irradiated brown dwarf WD 0137-349B}",
      journal = {\mnras},
         year = 2020,
        month = aug,
       volume = {496},
       number = {4},
        pages = {4674-4687},
          doi = {10.1093/mnras/staa1882},
archivePrefix = {arXiv},
       eprint = {2001.06558},
 primaryClass = {astro-ph.EP},
       adsurl = {https://ui.adsabs.harvard.edu/abs/2020MNRAS.496.4674L}
}

@ARTICLE{Lew2022,
       author = {{Lew}, Ben W.~P. and {Apai}, D{\'a}niel and {Zhou}, Yifan and {Marley}, Mark and {Mayorga}, L.~C. and {Tan}, Xianyu and {Parmentier}, Vivien and {Casewell}, Sarah and {Xu (许偲艺)}, Siyi},
        title = "{Mapping the Pressure-dependent Day-Night Temperature Contrast of a Strongly Irradiated Atmosphere with HST Spectroscopic Phase Curve}",
      journal = {\aj},
         year = 2022,
        month = jan,
       volume = {163},
       number = {1},
          eid = {8},
        pages = {8},
          doi = {10.3847/1538-3881/ac3001},
archivePrefix = {arXiv},
       eprint = {2110.10158},
 primaryClass = {astro-ph.SR},
       adsurl = {https://ui.adsabs.harvard.edu/abs/2022AJ....163....8L}
}

@article{komacek2016,
  title={Atmospheric circulation of hot Jupiters: dayside--nightside temperature differences},
  author={Komacek, Thaddeus D and Showman, Adam P},
  journal={The Astrophysical Journal},
  volume={821},
  number={1},
  pages={16},
  year={2016},
  doi={10.3847/0004-637X/821/1/16},
  publisher={IOP Publishing}
}

@article{parmentier2026,
  title={Horizontal transport as a source of disequilibrium chemistry on the nightside of a hot exoplanet},
  author={Parmentier, Vivien and Stevenson, Kevin B and Welbanks, Luis and Taylor, Jake and Schlawin, Everett and Coulombe, Louis-Philippe and Tang, Yao and Line, Mike and Shivkumar, Hinna and Tan, Xianyu and others},
  journal={Nature Astronomy},
  pages={1--9},
  year={2026},
  url={https://doi.org/10.1038/s41550-026-02845-2},
  publisher={Nature Publishing Group UK London}
}

@article{nasedkin2025,
  title={The JWST weather report: Retrieving temperature variations, auroral heating, and static cloud coverage on SIMP-0136},
  author={Nasedkin, Evert and Schrader, Merle and Vos, Johanna M and Biller, Beth and Burningham, Ben and Cowan, Nicolas B and Faherty, JK and Gonzales, Eileen and Lam, Madeline B and McCarthy, Allison M and others},
  journal={Astronomy \& Astrophysics},
  volume={702},
  pages={A1},
  year={2025},
  url={https://doi.org/10.1051/0004-6361/202555370},
  publisher={EDP Sciences}
}

@article{charnay2025,
  title={Latitudinal chemical and cloud variations in the atmosphere of a brown dwarf},
  author={Charnay, Benjamin and de Regt, Sam and Ravet, Mathieu and Teinturier, Lucas and Kiefer, Flavien and Chauvin, Ga{\"e}l and Denis, Allan and Bonnefoy, M and Palma-Bifani, Paulina and Radcliffe, Alice and others},
  DOI={https://doi.org/10.21203/rs.3.rs-7820642/v1},
  year={2025}
}

@article{showman2002,
  title={Atmospheric circulation and tides of “51 Pegasus b-like” planets},
  author={Showman, Adam P and Guillot, Tristan},
  journal={Astronomy \& Astrophysics},
  volume={385},
  number={1},
  pages={166--180},
  year={2002},
  doi={10.1051/0004-6361:20020101},
  publisher={EDP Sciences}
}

@article{helling2019,
  title={Exoplanet clouds},
  author={Helling, Christiane},
  journal={Annual Review of Earth and Planetary Sciences},
  volume={47},
  number={1},
  pages={583--606},
  year={2019},
  url={https://doi.org/10.1146/annurev-earth-053018060401},
  publisher={Annual Reviews}
}

@article{farihi2004,
  title={A possible brown dwarf companion to the white dwarf GD 1400},
  author={Farihi, Jay and Christopher, M},
  journal={The Astronomical Journal},
  volume={128},
  number={4},
  pages={1868--1871},
  DOI={10.1086/423919},
  year={2004}
}

@article{burleigh2006,
  title={A near-infrared spectroscopic detection of the brown dwarf in the post common envelope binary WD 0137-349},
  author={Burleigh, Matthew R and Hogan, E and Dobbie, PD and Napiwotzki, R and Maxted, PFL},
  journal={Monthly Notices of the Royal Astronomical Society: Letters},
  volume={373},
  number={1},
  pages={L55--L59},
  year={2006},
  url={https://doi.org/10.1111/j.1745-3933.2006.00242.x},
  publisher={Blackwell Science Ltd Oxford, UK}
}

@article{steele2013,
  title={NLTT 5306: the shortest period detached white dwarf+ brown dwarf binary},
  author={Steele, PR and Saglia, RP and Burleigh, Matthew R and Marsh, TR and G{\"a}nsicke, BT and Lawrie, K and Cappetta, M and Girven, Jonathan and Napiwotzki, R},
  journal={Monthly Notices of the Royal Astronomical Society},
  volume={429},
  number={4},
  pages={3492--3500},
  year={2013},
  doi={https://doi.org/10.1093/mnras/sts620},
  publisher={Oxford University Press}
}

@article{casewell2012,
  title={WD0837+ 185: the formation and evolution of an extreme mass-ratio white-dwarf--brown-dwarf binary in Praesepe},
  author={Casewell, Sarah L and Burleigh, MR and Wynn, GA and Alexander, RD and Napiwotzki, R and Lawrie, KA and Dobbie, PD and Jameson, RF and Hodgkin, ST},
  journal={The Astrophysical Journal Letters},
  volume={759},
  number={2},
  pages={L34},
  year={2012},
  DOI={10.1088/2041-8205/759/2/L34},
  publisher={The American Astronomical Society}
}

@article{gao2018,
  title={Microphysics of KCl and ZnS Clouds on GJ 1214 b},
  author={Gao, Peter and Benneke, Bj{\"o}rn},
  journal={The Astrophysical Journal},
  volume={863},
  number={2},
  pages={165},
  year={2018},
  DOI={10.3847/1538-4357/aad461},
  publisher={The American Astronomical Society}
}

@article{parmentier2021,
  title={The cloudy shape of hot Jupiter thermal phase curves},
  author={Parmentier, Vivien and Showman, Adam P and Fortney, Jonathan J},
  journal={Monthly Notices of the Royal Astronomical Society},
  volume={501},
  number={1},
  pages={78--108},
  year={2021},
  doi={10.1093/mnras/staa3418},
  publisher={Oxford University Press}
}

@article{vos2017,
  title={The viewing geometry of brown dwarfs influences their observed colors and variability amplitudes},
  author={Vos, Johanna M and Allers, Katelyn N and Biller, Beth A},
  journal={The Astrophysical Journal},
  volume={842},
  number={2},
  pages={78},
  year={2017},
  url={https://doi.org/10.3847/1538-4357/aa73cf},
  publisher={The American Astronomical Society}
}

@article{tan2021a,
  title={Atmospheric circulation of brown dwarfs and directly imaged exoplanets driven by cloud radiative feedback: global and equatorial dynamics},
  author={Tan, Xianyu and Showman, Adam P},
  journal={Monthly Notices of the Royal Astronomical Society},
  volume={502},
  number={2},
  pages={2198--2219},
  year={2021},
  url={https://doi.org/10.1093/mnras/stab097},
  publisher={Oxford University Press}
}

@article{tan2021b,
  title={Atmospheric circulation of brown dwarfs and directly imaged exoplanets driven by cloud radiative feedback: effects of rotation},
  author={Tan, Xianyu and Showman, Adam P},
  journal={Monthly Notices of the Royal Astronomical Society},
  volume={502},
  number={1},
  pages={678--699},
  year={2021},
  url={https://doi.org/10.1093/mnras/stab060},
  publisher={Oxford University Press}
}

@article{tan2019,
  title={Atmospheric variability driven by radiative cloud feedback in brown dwarfs and directly imaged extrasolar giant planets},
  author={Tan, Xianyu and Showman, Adam P},
  journal={The Astrophysical Journal},
  volume={874},
  number={2},
  pages={111},
  year={2019},
  DOI={10.3847/1538-4357/ab0c07},
  publisher={IOP Publishing}
}

@article{morley2012,
  title={Neglected clouds in T and Y dwarf atmospheres},
  author={Morley, Caroline V and Fortney, Jonathan J and Marley, Mark S and Visscher, Channon and Saumon, Didier and Leggett, SK},
  journal={The Astrophysical Journal},
  volume={756},
  number={2},
  pages={172},
  year={2012},
  doi={10.1088/0004-637X/756/2/172},
  publisher={IOP Publishing}
}

@book{Vallis2006,
author={Vallis, Geoffrey K},
title = {Atmospheric and Oceanic Fluid Dynamics},
publisher = {Cambridge University Press},
address = {Cambridge, U.K.},
year = {2006},
volume={770},
pages = {745},
}

@article{batalha2020,
  title={Refractive Indices For Virga Exoplanet Cloud Model},
  author={Batalha, Natasha and Marley, Mark},
  journal={Zenodo, doi},
  url={https://zenodo.org/records/15886530},
  volume={10},
  year={2020}
}

@article{montaner1979,
  title={Optical constants of sodium sulphide},
  author={Montaner, A and Galtier, M and Benoit, C and Bill, H},
  year={1979}
}

@article{khachai2009,
  title={FP-APW+ lo calculations of the electronic and optical properties of alkali metal sulfides under pressure},
  author={Khachai, H and Khenata, R and Bouhemadou, A and Haddou, A and Reshak, Ali H and Amrani, B and Rached, D and Soudini, B},
  journal={Journal of Physics: Condensed Matter},
  volume={21},
  number={9},
  pages={095404},
  DOI={10.1088/0953-8984/21/9/095404},
  year={2009}
}

@article{jager2003,
  title={Steps toward interstellar silicate mineralogy: VII. Spectral properties and crystallization behaviour of magnesium silicates produced by the sol-gel method},
  author={J{\"a}ger, Cornelia and Dorschner, J and Mutschke, Harald and Posch, Th and Henning, Th},
  journal={Astronomy \& Astrophysics},
  volume={408},
  number={1},
  pages={193--204},
  year={2003},
  url={https://doi.org/10.1051/0004-6361:20030916},
  publisher={EDP Sciences}
}

@incollection{palik1997,
  title={Iron Pyrite (FeS2)},
  author={Palik, ED},
  booktitle={Handbook of Optical Constants of Solids},
  pages={507--517},
  year={1997},
  doi={10.1016/B978-012544415-6.50115-1},
  publisher={Elsevier}
}

@article{Kirkpatrick2012,
  title={Further defining spectral type “Y” and exploring the low-mass end of the field brown dwarf mass function},
  author={Davy Kirkpatrick, J and Gelino, Christopher R and Cushing, Michael C and Mace, Gregory N and Griffith, Roger L and Skrutskie, Michael F and Marsh, Kenneth A and Wright, Edward L and Eisenhardt, Peter R and McLean, Ian S and others},
  journal={The Astrophysical Journal},
  volume={753},
  number={2},
  pages={156},
  year={2012},
  DOI={10.1088/0004-637X/753/2/156},
  publisher={The American Astronomical Society}
}

@article{leggett2012,
  title={THE PROPERTIES OF THE 500 K DWARF UGPS J072227. 51- 054031.2 AND A STUDY OF THE FAR-RED FLUX OF COLD BROWN DWARFS},
  author={Leggett, SK and Saumon, D and Marley, MS and Lodders, K and Canty, J and Lucas, Philippe and Smart, RL and Tinney, CG and Homeier, D and Allard, F and others},
  journal={The Astrophysical Journal},
  volume={748},
  number={2},
  pages={74},
  year={2012},
  DOI={10.1088/0004-637X/748/2/74},
  publisher={The American Astronomical Society}
}

@article{adams2022,
  title={Spatially resolved modeling of optical albedos for a sample of six hot Jupiters},
  author={Adams, Danica J and Kataria, Tiffany and Batalha, Natasha E and Gao, Peter and Knutson, Heather A},
  journal={The Astrophysical Journal},
  volume={926},
  number={2},
  pages={157},
  year={2022},
  DOI={10.3847/1538-4357/ac3d32},
  publisher={The American Astronomical Society}
}

@article{robbins2022,
  title={Cloudy and cloud-free thermal phase curves with PICASO: Applications to WASP-43b},
  author={Robbins-Blanch, Nina and Kataria, Tiffany and Batalha, Natasha E and Adams, Danica J},
  journal={The Astrophysical Journal},
  volume={930},
  number={1},
  pages={93},
  year={2022},
  DOI={10.3847/1538-4357/ac658c},
  publisher={The American Astronomical Society}
}

@article{menou2020,
  title={Hot Jupiter atmospheric flows at high resolution},
  author={Menou, Kristen},
  journal={Monthly Notices of the Royal Astronomical Society},
  volume={493},
  number={4},
  pages={5038--5044},
  year={2020},
  url={https://doi.org/10.1093/mnras/staa532},
  publisher={Oxford University Press}
}

@article{skinner2021,
  title={Numerical convergence of hot-Jupiter atmospheric flow solutions},
  author={Skinner, J W and Cho, J YK},
  journal={Monthly Notices of the Royal Astronomical Society},
  volume={504},
  number={4},
  pages={5172--5187},
  year={2021},
  url={https://doi.org/10.1093/mnras/stab971},
  publisher={Oxford University Press}
}

@book{vallis2017atmospheric,
  title={Atmospheric and oceanic fluid dynamics},
  author={Vallis, Geoffrey K},
  year={2017},
  publisher={Cambridge University Press}
}

@article{broskilaing2026,
  title={Asymmetric Nightside CO2 Features, Inefficient Heat Transport, and Precise Evolutionary Constraints: Spectroscopic Phase Curves Reveal the Past and Present of a White Dwarf--Brown Dwarf Binary},
  author={Broski-Laing, Daphne and Zhou, Yifan and Lothringer, Joshua D and Apai, Daniel and French, Jenni R and Casewell, Sarah L and Mayorga, LC and Shin, Lael and Lew, Ben WP and Tan, Xianyu and others},
  journal={The Astrophysical Journal},
  volume={1009},
  number={2},
  pages={177},
  year={2026},
  DOI={10.3847/1538-4357/ae9940},
  publisher={The American Astronomical Society}
}
\bibliographystyle{aasjournal}

\end{document}